\documentclass[onecolumn,showkeys,aps,prd]{revtex4}

\usepackage{amsfonts}
\usepackage{latexsym}
\usepackage{amsmath,amssymb}
\usepackage{mathtools}
\usepackage{lineno}
\usepackage{bbm}

\newcommand{\sout}[1]{}
\newcommand{\rsout}[1]{}
\newcommand{\bsout}[1]{}
\newcommand{\red}[1]{#1}

\newcommand{\esout}[1]{}

\renewcommand{\dj}{d\kern-0.4em\char"16\kern-0.1em}
\newcommand{\Dj}{\mbox{\raise0.3ex\hbox{-}\kern-0.4em D}}

\newcommand{\lc}{\varepsilon}

\newcommand{\nablar}{\stackrel{\rightarrow}{\nabla}\!\!{}}
\newcommand{\nablal}{\stackrel{\leftarrow}{\nabla}\!\!{}}
\newcommand{\nablalr}{\stackrel{\leftrightarrow}{\nabla}\!\!{}}
\newcommand{\ds}{\displaystyle}

\newcommand{\diag}{\mathop{\rm diag}\nolimits}

\newcommand{\rmd}{\mathrm{d}}

\newcommand{\realni}{\ensuremath{\mathbb{R}}}
\newcommand{\kompleksni}{\ensuremath{\mathbb{C}}}

\newcommand{\cC}{{\cal C}}

\newcommand{\cF}{{\cal F}}
\newcommand{\cG}{{\cal G}}
\newcommand{\cH}{{\cal H}}

\newcommand{\cM}{{\cal M}}

\newcommand{\cO}{{\cal O}}

\newcommand{\one}{\ensuremath{\mathbbm{1}}}

\newcommand{\dual}{\text{\boldmath$\,\star$}}

\begin{document}

\title{Symmetry breaking mechanisms of the $3BF$ action \\ for the Standard Model coupled to gravity}

\author{Pavle Stipsi\'c}
 \email{pstipsic@ipb.ac.rs}
\affiliation{Institute of Physics, University of Belgrade, Pregrevica 118, 11080 Belgrade, Serbia}

\author{Marko Vojinovi\'c}
 \email{vmarko@ipb.ac.rs}
\affiliation{Institute of Physics, University of Belgrade, Pregrevica 118, 11080 Belgrade, Serbia}


\keywords{quantum gravity, higher gauge theory, $3$-group, $3BF$ action, symmetry breaking, Higgs mechanism}

\begin{abstract}
We study the details of the explicit and spontaneous symmetry breaking of the constrained $3BF$ action representing the Standard Model coupled to Einstein-Cartan gravity. First we discuss how each particular constraint breaks the original symmetry of the topological $3BF$ action. Then we investigate the spontaneous symmetry breaking and the Higgs mechanism for the electroweak theory in the constrained $3BF$ form, in order to demonstrate that they can indeed be performed in the framework of higher gauge theory. A formulation of the Proca action as a constrained $3BF$ theory is also studied in detail.
\end{abstract}

\maketitle

\section{\label{SecI}Introduction}

The formulation of a quantum theory of the gravitational field represents one of the main open problems in the modern fundamental theoretical physics. Over the years, there have been many attempts to tackle this problem, and several major approaches have been developed, including String Theory (ST) \cite{Polchinski1,Polchinski2}, Loop Quantum Gravity (LQG) \cite{Rovelli2004,Thiemann}, and various other frameworks. Each of these approaches has its own set of advantages and disadvantages. In particular, the covariant version of the LQG approach \cite{RovelliVidotto2014} focuses on the rigorous definition of the gravitational path integral, which is used as a key ingredient in defining the corresponding quantum theory. One of the main advantages of the covariant LQG lies in the fact that such a rigorous definition can in fact be formulated, using the so-called {\em spinfoam quantization procedure}, and the gravitational field can be quantized successfully. On the other hand, the main disadvantage lies in the fact that the spinfoam quantization procedure works well for the pure gravitational field, but is not compatible with all other fields in nature (collectively called matter fields) \cite{SpinfoamFermions,VojinovicCosineProblem,MikovicVojinovicBook}. In other words, while it is possible to quantize the gravitational field, it is not straightforward to quantize the gravitational field with matter.

In recent years, progress has been made to circumvent this disadvantage. One of the promising avenues is based on the so-called {\em higher gauge theory} \cite{BaezHuerta,Crane2003}, which represents a framework that generalizes the notion of symmetry using mathematical techniques of higher category theory. Attention is mostly focused on the categorical structures called $n$-groups, which are a certain type of generalization of the algebraic notion of a group, and are used instead of groups to describe the gauge symmetry of the theory \cite{MikovicVojinovic2012} (for various other applications of $n$-groups to physics see for example \cite{Li2019,SaemannWolf2014,JurcoEtal2005,HNY2020,HNY2021,SaemannWolf2014b,SWY2021a,SWY2021b,HNY2021b,HNY2021c,JurcoSaemannWolf2016,SaemannWolf2017,JurcoEtal2019b}). In particular, the structure of $3$-groups appears to be suitable to give an algebraic description of all relevant fields in nature --- the gravitational field, the Yang-Mills field, the scalar field and the Dirac field \cite{Radenkovic2019}. On the other hand, the structure of the $3$-group lends itself nicely to a generalization of the spinfoam quantization procedure \cite{Radenkovic2022b}, which opens the door to study the quantization of the gravitational field with matter within a unified mathematical description.

\red{One of the central elements in the above construction is a notion of the $BF$ theory and its higher gauge theory generalizations called $nBF$ theories. Historically, one of the natural approaches to these theories relies on the Batalin-Vilkovisky formalism \cite{BV1,BV2,BV3,BV4,BV5}, and gives rise to the so-called Alexandrov-Kontsevich-Schwarz-Zaboronsky (AKSZ) construction, see \cite{hep-th/9502010} and further developments in \cite{MaximParent,MaximPresymplectic,MaximPresymplecticMinimal,Alberto(Maxim),AKSZ-Manin(Maxim),Jan(Maxim)}. The classical formulation of general relativity and other theories of gravity based on the $BF$ theory have initial results in the work of Plebanski \cite{Plebanski}, see also \cite{baez2000,BFgravity2016,Baez1996} for a comprehensive review of various models. The 2-group formulation, called $2BF$ or $BFCG$ model, was first introduced in \cite{GirelliPfeifferPopescu2008,FariaMartinsMikovic2011} and further studied in \cite{MikovicOliveira2014,Mikovic2015,MOV2016,MOV2019,Asante2020,Girelli2021}. The classical $3BF$ and $4BF$ theories were formulated in \cite{Radenkovic2019} and \cite{MV2020}, respectively. At the quantum level, $nBF$ theories give rise to a class of topological quantum field theories, first introduced in the works of Porter \cite{Porter}, see also \cite{Radenkovic2022b,Radenkovic2024}.
}

The higher gauge theory programme based on $3$-groups has recently given some promising concrete results. First, it was understood how to construct a classical action that describes the full Standard Model (SM) naturally coupled to Einstein-Cartan gravity \red{(our naming convention follows the textbook \cite{BlagojevicBook})}, so that it has a form compatible with the generalized spinfoam quantization procedure \cite{Radenkovic2019}. This amounts to the reformulation of the classical theory into a form of the so-called constrained $3BF$ action. Such an action consists of two main parts --- the topological $3BF$ part, specified by the postulated structure of a $3$-group, and the constraint part, which deforms the topological theory into a non-topological one, with nontrivial dynamics. \rsout{A lot of research has also been focused on simpler models based on $BF$ and $2BF$ theories, see for example}\esout{ \cite{baez2000,BFgravity2016,GirelliPfeifferPopescu2008,FariaMartinsMikovic2011,MikovicOliveira2014,Mikovic2015,MOV2016,MOV2019,Asante2020,Girelli2021}}\rsout{, and there is even one model based on a $4BF$ theory}\esout{ \cite{MV2020}. }Next, the quantization procedure for the topological sector has been sucessfully implemented, leading to a formulation of the path integral corresponding to the topological quantum field theory (TQFT) based on a given $3$-group \cite{Radenkovic2022b}. In addition to these results, the symmetries of the topological $3BF$ theory have been studied in full detail \cite{Radenkovic2020,Radenkovic2022a}, leading to deeper understanding of the various properties of the models. Some important mathematical results have also been established \cite{SaemannWolf2014b,martins2011,Wang2014}. Nevertheless, the symmetries of the constrained $3BF$ action, which represents a realistic classical theory, have not been studied so far. The main purpose of this paper is to fill this missing step, and study the symmetries of the constrained $3BF$ action.

From the structure of the constrained $3BF$ action, it is straightforward to see that the topological sector has a certain (large) symmetry, while the constraints mainly break this symmetry to one of its subgroups. Therefore, our work focuses on the study of various symmetry breaking mechanisms and the role played by each individual constraint. In particular, we examine precisely how each constraint individually breaks the $3BF$ symmetry group and down to which subgroup. As it turns out, some constraints have bigger influence and break the symmetry down to a smaller subgroup, while other constraints have smaller influence and break the symmetry only slightly. We shall also find out that one part of the symmetry group remains unbroken even in the presence of all constraints. All these results are then organized and presented in a form of a table. After the analysis of the explicit symmetry breaking, our attention turns to the details of the spontaneous symmetry breaking, and the Higgs mechanism. This is very important, since the Higgs mechanism is crucial for the Standard Model, and it is not clear whether the constrained $3BF$ formulation of the SM action admits the implementation of the Higgs mechanism with the same outcome as the ordinary textbook formulation of the theory. It turns out that the Higgs mechanism does indeed yield the expected outcome, but the details of the implementation of spontaneous symmetry breaking are very far from the ordinary textbook approach. Instead, a set of completely new calculational techniques has been developed, including one theorem, and they represent our second main result. These new techniques are necessary, due to the fact that the constrained $3BF$ formulation of the SM action is based on a very different set of variables, compared to the textbook SM Lagrangian. Finally, as one of the major steps in this analysis, we also provide a formulation of the Proca action within the framework of higher gauge theory, and discuss explicitly three different versions of the constrained $3BF$ action for the Proca theory coupled to gravity. This also represents a novel result, not present in the previous literature.

The layout of the paper is as follows. In Section \ref{secII}, we present a short overview of the higher gauge theory construction of the SM action coupled to gravity. We introduce the notion of a $3$-group and the corresponding topological $3BF$ action, and then we demonstrate how these should be chosen and deformed with constraint terms in order to reproduce the Standard Model coupled to Einstein-Cartan gravity. Section \ref{secIII} contains the short review of the gauge symmetry group of the topological $3BF$ action, and the analysis how this group is being broken by each individual constraint term in the action. This represents the study of the explicit symmetry breaking as a consequence of the constraints present in the theory. Section \ref{secIV} is devoted to the $3$-group formulation of the Proca action. The Proca action is a necessary ingredient one needs to understand, in order to compare it to the action obtained via spontaneous symmetry breaking. The analysis of the spontaneous symmetry breaking and the Higgs mechanism is presented in full detail in Section \ref{secV}, discussing the most convenient example of electroweak theory. Despite the fact that it is conceptually the same as the ordinary Higgs mechanism described in textbooks, the specific properties of the $3BF$ formulation of the action renders the technical details of the procedure highly nontrivial, and represents one of the main results of the paper. Our concluding remarks are given in Section \ref{secVI}, with a summary and a discussion of the results. The Appendices contain some additional technical material.

Our notation and conventions are as follows. Spacetime indices, denoted by the mid-alphabet Greek letters $\mu,\nu,\dots$, are raised and lowered by the spacetime metric $g_{\mu\nu}$. The Lorentz metric is denoted as $\eta_{ab} = \diag (-1,+1,+1,+1)$. The indices that are counting the generators of Lie groups $G$, $H$, and $L$ are denoted with initial Greek letters $\alpha, \beta, \dots$, lowercase initial Latin letters $a, b, c,\dots$, and uppercase Latin indices $A,B,C,\dots$, respectively. The generators themselves are typically denoted as $\tau_\alpha$, $t_a$ and $T_A$, respectively. We work in the natural system of units, defined by $c=\hbar=1$ and $G = l_p^2$, where $l_p$ is the Planck length. All additional notation and conventions used throughout the paper are explicitly defined in the text where they first appear.

\section{\label{secII}Review of the $3BF$ theory with constraints}

The main idea of the so-called higher gauge theory approach \cite{BaezHuerta} is to describe the symmetry of a physical theory with a mathematical structure that is different from an ordinary Lie group. In particular, in the context of category theory description of groups, the natural generalizations are the structures called $n$-groups. \red{A reader interested in the mathematical details of higher category theory, $n$-groups and $L_\infty$ algebras is referred to corresponing literature \cite{BaezHuerta,BaezDolan1995,Li2019,Crane2003,JurcoEtal2005,Conduche1984,Yetter1993,Porter,GirreliPfeiffer2004,GirelliPfeifferPopescu2008,FariaMartinsMikovic2011,martins2011,Wang2014,Berends1985,LadaStasheff1993,Stasheff1998,HohmZwiebach2017,JurcoEtal2019b}.} For the purpose of this work, the attention focuses on the notion of a strict Lie $3$-group structure as a symmetry of the theory.

A strict Lie $3$-group is defined as a $3$-category over a single object with invertibile $1$-, $2$- and $3$-morphisms, and it is equivalent to a Lie $2$-crossed module $(L\xrightarrow{\delta}H\xrightarrow{\partial}G,\ \triangleright,\ \{\_\, , \_\}_{\rm pf})$. A $2$-crossed module consists of three Lie groups $L$, $H$, and $G$, with the homomorphism $\partial : H \to G$, the homomorphism $\delta:L\to H$, the action $\triangleright: G\times X \to X$ (where $X\in \{ G, H, L \}$), and Peiffer lifting $\{\_\, , \_\}_{\rm pf}: H\times H \to L$. In order to give rise to a Lie $2$-crossed module, all these maps must satisfy a set of appropriate axioms \cite{Radenkovic2019}.

One of the main benefits of the categorical generalization of the notion of a Lie group lies in the corresponding generalization of certain notions in differential geometry. In particular, for the case of a Lie $3$-group, one can generalize the notion of parallel transport along a curve to the notions of parallel transport across a surface and through a volume. These operations are described by the so-called $2$- and $3$-holonomies, which are in turn constructed from the $\mathfrak{h}$-valued connection two-form $\beta$ and the $\mathfrak{l}$-valued connection three-form $\gamma$, in addition to the ordinary $\mathfrak{g}$-valued connection one-form $\alpha$. The triple $(\alpha,\beta,\gamma)$ is called a $3$-connection, and $\mathfrak{g}$, $\mathfrak{h}$, $\mathfrak{l}$ are Lie algebras of the Lie groups $G$, $H$, and $L$, respectively.

The mathematical structure of a $3$-group gives rise to a natural choice of an action, called $3BF$ action, that can be constructed from the $3$-connection. The $3BF$ action is purely topological, and defined as:
\begin{equation} \label{eq:3BFaction}
S_{3BF}=\int_{\cM_4} \langle B\wedge {\cal F}\rangle_\mathfrak{g}+\langle C\wedge {\cal G}\rangle_\mathfrak{h}+\langle D\wedge {\cal H}\rangle_\mathfrak{l}.
\end{equation} 
Quantities $B$, $C$ and $D$ are two-, one- and zero-forms, which play the role of the Lagrange multipliers, and they are elements of algebras $\mathfrak{g}$, $\mathfrak{h}$ and $\mathfrak{l}$, respectively. The field strengths $\cF$, $\cG$ and $\cH$ are defined as
\begin{equation} \label{eq:ThreeCurvatureDef}
{\cal F}={\rm d}\alpha+\alpha\wedge \alpha-\partial\beta\,,\qquad {\cal G}={\rm d}\beta+\alpha\wedge^{\triangleright}\beta-\delta\gamma\,,\qquad
{\cal H}={\rm d}\gamma+\alpha\wedge^{\triangleright}\gamma+\{\beta\wedge\beta\}_{\rm pf}\,,
\end{equation}
and they are called fake curvatures for the connection one-form $\alpha$, two-form $\beta$ and three-form $\gamma$. Bilinear forms $\langle\_\, ,\_\rangle_{\mathfrak{g}}$, $\langle\_\, ,\_\rangle_{\mathfrak{h}}$  and $\langle\_\, ,\_\rangle_{\mathfrak{l}}$  are assumed to be symmetric, nondegenerate and $G$-invariant, and they map a pair of algebra elements into a real number. Evaluated on the corresponding basis vectors, the bilinear forms are written in components as follows:
\begin{equation}
\langle \tau_\alpha\, ,\tau_\beta\rangle_{\mathfrak{g}} = g_{\alpha\beta} \,, \qquad
\langle t_a\, ,t_b\rangle_{\mathfrak{h}} = g_{ab} \,, \qquad
\langle T_A\, ,T_B\rangle_{\mathfrak{l}} = g_{AB} \,.
\end{equation}
Since the bilinear forms are assumed to be nondegenerate, the inverses of the above components also exist, denoted as $g^{\alpha\beta}$, $g^{ab}$ and $g^{AB}$. They are collectively used to raise and lower all group indices as necessary.

Varying the action (\ref{eq:3BFaction}) with respect to Lagrange multipliers, one obtains the equations of motion for fake curvatures:
\begin{equation}
{\cal F}=0\,,\qquad
{\cal G}=0\,,\qquad
{\cal H}=0\,.
\end{equation}
Also, varying with respect to the connections $\alpha$, $\beta$ and $\gamma$, we get the remaining three equations of motion:
\begin{eqnarray}
\nabla B_{\alpha}-\triangleright_{\alpha a}{}^{b}C_b\wedge \beta^a+\triangleright_{\alpha B}{}^{A}D_A\wedge \gamma^B&=&0\,,\\
\nabla C_a-\partial_a{}^{\alpha}B_{\alpha}+2X_{(ab)}{}^{A}D_A\wedge\beta^b&=&0\,,\\
\nabla D_A+\delta_A{}^{a}C_a&=&0\,.
\end{eqnarray}
Here the quantities $X_{ab}{}^A$ are components of the Peiffer lifting evaluated on the basis, $\{t_a , t_b\}_{\rm pf} \equiv X_{ab}{}^A T_A$. An analogous notation is used for the homomorphisms $\partial$ and $\delta$, and the action $\triangleright$:
\begin{equation}
\partial t_a = \partial_a{}^\alpha \tau_\alpha \,, \qquad
\delta T_A = \delta_A{}^a t_a\,, \qquad
\tau_\alpha \triangleright t_a = \triangleright_{\alpha a}{}^b t_b \,, \qquad
\tau_\alpha \triangleright T_A = \triangleright_{\alpha A}{}^B T_B \,.
\end{equation}

At this point it is important to note one feature of the relationship between a 3-group and the three bilinear forms. Specifically, the requirement that the bilinear forms must be $G$-invariant places a restriction on the allowed choice of the action $\triangleright$. This is specified in the following theorem.

\medskip

\textbf{Theorem.} Given a $2$-crossed module $(L\xrightarrow{\delta}H\xrightarrow{\partial}G,\ \triangleright,\ \{\_\, , \_\}_{\rm pf})$ and symmetric, nondegenerate bilinear forms $\langle\_\, ,\_\rangle_{\mathfrak{g}}$, $\langle\_\, ,\_\rangle_{\mathfrak{h}}$  and $\langle\_\, ,\_\rangle_{\mathfrak{l}}$, if the bilinear forms are $G$-invariant then the components of the action $\triangleright_{\alpha\beta\gamma}$, $\triangleright_{\alpha ab}$ and $\triangleright_{\alpha AB}$ are antisymmetric with respect to the second and third index. In addition, there exists a choice of basis in Lie algebras $\mathfrak{g}$, $\mathfrak{h}$ and $\mathfrak{l}$ such that $\triangleright_{\alpha\beta}{}^\gamma$, $\triangleright_{\alpha a}{}^b$ and $\triangleright_{\alpha A}{}^B$ have vanishing diagonal elements with respect to the second and third index, and in this basis the bilinear form is also diagonal.

\medskip

For the proof of the Theorem, see Appendix \ref{app:a}. It is important to stress that these restrictions on the action $\triangleright$ arise only due to assumed $G$-invariance of the bilinear forms, and do not hold otherwise. Furthermore, these restrictions will play an important role later in Section \ref{secV}, in the discussion of the Higgs mechanism.

\red{As noted in the Introduction section, one can apply $BF$ and more generally $nBF$ actions to construct physically interesting models and even realistic theories. This is typically performed by adding an additional term to the topological action, called a constraint term, which deforms the theory and may give rise to physical degrees of freedom. In the context of ordinary $BF$ theory, relevant physical models include Plebanski formulation of general relativity, Husain-Kucha\v r model, MacDowell-Mansouri model, JT model, and general relativity in $2+1$ dimensions and more than $4$ dimensions. A comprehensive reivew of these models is given in \cite{BFgravity2016}. Regarding the $2BF$ theory, relevant models include Einstein-Cartan gravity and Yang-Mills theory, see \cite{MikovicVojinovic2012,Radenkovic2019}. In the context of $3BF$ theory, one can construct models with matter fields in addition to gravity and Yang-Mills. In particular, models with scalar, Dirac, Weyl and Majorana fields coupled to Einstein-Cartan gravity and Yang-Mills theory have been constructed, including the full Standard Model coupled to Einstein-Cartan gravity. For a review of these models see \cite{Radenkovic2019}. Finally, there is also a similar construction based on $4BF$ model, see \cite{MV2020}.

In this work, we will focus our attention on a constrained $3BF$ action giving rise to the Einstein-Cartan gravity coupled to the full Standard Model, and the electroweak model, as well as a formulation of the Proca theory.}

In order to construct a physical theory based on a $3BF$ action, we need to introduce some constraints between the fields and choose an appropriate $3$-group as a gauge symmetry structure. The constraints are discussed below, while the choice of the $3$-group is as follows. A simple $3$-group which corresponds to the Standard Model of elementary particles coupled to Einstein-Cartan gravity in the usual way is called the Standard Model $3$-group \cite{Radenkovic2019}, and is defined by the following choice:
\begin{equation}
G = SO(3,1)\times SU(3)\times SU(2) \times U(1) \,, \qquad H = \mathbb{R}^4\,, \qquad L = \mathbb{C}^4\times\mathbb{G}^{64}\times\mathbb{G}^{64}\times\mathbb{G}^{64}\,.
\end{equation}
The group $G$ is a product of the Lorentz group and the usual internal gauge symmetry groups of the Standard Model. The group $H$ represents spacetime translations. The choice of the group $L$ corresponds to the Higgs and fermion sector of the Standard Model, where $\mathbb{G}$ denotes the Grassmann algebra. In addition to this choice of groups, we choose the maps $\partial$, $\delta$ and $\{\_\ , \_\}_{\rm pf}$ to be trivial:
\begin{equation}
\partial h = \one_G \,, \qquad \delta l = \one_H \,, \qquad \{h_1 , h_2\}_{\rm pf} = \one_L\,,
\end{equation}
for every $h,h_1,h_2 \in H$ and $l \in L$.

In order to complete the definition of the $3$-group, we also choose the map $\triangleright$ as follows. Because of the specific structure of the group $G$, it is natural to distinguish $\mathfrak{g}$-indices for the Lorentz part from the internal part, and we will write the former in pairs of small alphabet indices in angular brackets, $[ab]$, while the latter will remain denoted with Greek letters from the beginning of the alphabet. Denoting the structure constants for the internal subgroup $SU(3)\times SU(2) \times U(1)$ as $f_{\alpha\beta}{}^{\gamma}$, the action of the group $G$ on itself is defined as:
\begin{equation} \label{eq:actionOfLorentzPartOfG}
\triangleright_{[ab][cd]}{}^{[ef]}\equiv f_{[ab][cd]}{}^{[ef]}=\frac{1}{2}\left(\eta_{[a|c}\delta_{|b]}^{[f|}\delta_{d}^{|e]}-\eta_{[a|d}\delta_{|b]}^{[f|}\delta_{c}^{|e]}\right)\,, \qquad
\triangleright_{[ab] \beta}{}^{\gamma}=0\,,
\end{equation}
\begin{equation} \label{eq:actionOfInternalPartOfG}
\triangleright_{\alpha \beta}{}^{\gamma} = f_{\alpha\beta}{}^{\gamma} \,, \qquad
\triangleright_{\alpha [ab]}{}^{[cd]}=0\,.
\end{equation}
Equations (\ref{eq:actionOfLorentzPartOfG}) define the action of the Lorentz subgroup on $G$, while equations (\ref{eq:actionOfInternalPartOfG}) define the action of the internal subgroup on $G$. The action of the Lorentz and internal subgroups of $G$ on the group $H$ is defined as:
\begin{equation}\label{eq:actionGOnH}
\triangleright_{[cd]a}{}^{b}=\frac{1}{2}\eta_{a[d|}\delta_{|c]}^b\,, \qquad
\triangleright_{\alpha a}{}^{b}=0\,.
\end{equation}
Finally, the action of the Lorentz and internal subgroups of $G$ on $L$ is given in a natural way, in accordance with the transformation properties of various fermions and the Higgs scalar. For example, the action of $G$ on left-isospin fermions is given as:
\begin{equation}
\triangleright_{[cd]A}{}^{B}=\frac{1}{2}\left(\sigma_{[cd]}\right)_A{}^{B}\,, \qquad
\triangleright_{\alpha A}{}^{B}=\frac{1}{2}\left(\sigma_{\alpha}\right)_A{}^{B}\,.
\end{equation}
Here the matrices $\left(\sigma_{\alpha}\right)_A{}^{B}$ are Pauli matrices, and $\left(\sigma_{[ab]}\right)_A{}^{B}=\frac{1}{4}[\gamma_a,\gamma_b]_A{}^{B}$, where $\gamma_a$ are the standard Dirac matrices satisfying the anticommutation rule $ \gamma_a\gamma_b + \gamma_b\gamma_a = -2 \eta_{ab}$. Here we also introduce $\gamma_5 \equiv - \gamma_0 \gamma_1 \gamma_2 \gamma_3$. In a similar way, one defines the action of group $G$ for all other fermions and scalars in the group $L$, depending on their precise transformation properties (see \cite{Radenkovic2019} for details).

In addition to the specification of the $3$-group, the action (\ref{eq:3BFaction}) also depends on the choice of bilinear forms. For the non-Abelian groups one can naturally choose the Cartan-Killing form, while for the Abelian groups there is no natural choice, and one is mostly restricted by the property that the bilinear form must be $G$-invariant. Taking this into account, for the Standard Model $3$-group and the action (\ref{eq:3BFaction}) we choose the bilinear forms as follows. For the algebra $\mathfrak{g}$, we have
\begin{equation}\label{snbfg}
g_{[ab][cd]}=\frac{1}{2}\eta_{d[a|}\eta_{|b]c}\,, \qquad
g_{\alpha\beta}=\delta_{\alpha\beta}\,, \qquad
g_{\alpha [ab]}=0\,.
\end{equation}
For the algebra $\mathfrak{h}$, due to the $G$-invariance, we have
\begin{equation}\label{snbfh}
g_{ab}=\eta_{ab}\,.
\end{equation}
Finally, for the algebra $\mathfrak{l}$ the situation is more complicated, since the Grassmann numbers anticommute. Namely, note that for general $A,B\in\mathfrak{l}$, we can write
\begin{equation}
\langle A,B \rangle_\mathfrak{l} = A^I B^J g_{IJ}\,, \qquad \langle B,A \rangle_\mathfrak{l} = B^J A^I g_{JI}\,.
\end{equation}
Since the bilinear form must be symmetric, the two expressions must be equal. However, depending on whether the coefficients $A^I$ and $B^J$ are Grassmann numbers or ordinary real numbers, they will either anticommute or commute, and consequently the component matrix $g_{IJ}$ of the bilinear form must be antisymmetric or symmetric, respectively. In our case, the generators $T_A$ of the algebra $\mathfrak{l}$ can be grouped into three classes: $T_{\tilde{A}}$ which belong to the Higgs sector, and a pair $T_{\hat{A}}$, $T^{\hat{A}}$ which belong to the fermion sector. Then, the components of the bilinear form can be written as
\begin{equation}
g_{AB} = \left[
\begin{array}{c|cc}
 \delta_{\tilde{A}\tilde{B}} & 0 & 0 \\ \hline
 0 & 0 & \delta_{\hat{A}}^{\hat{B}} \\
 0 & -\delta_{\hat{A}}^{\hat{B}} & 0 \\
\end{array}
\right]\,.
\end{equation}
The upper-left block corresponds to the algebra $\mathbb{C}^4$, while the bottom-right block corresponds to the algebra $\mathbb{G}^{64}\times\mathbb{G}^{64}\times\mathbb{G}^{64}$.

Once we have specified both the $3$-group and the bilinear forms, we can introduce the full classical action corresponding to the Standard Model coupled to Einstein-Cartan gravity. This action is written as the $3BF$ action (\ref{eq:3BFaction}) plus the constraint terms that give rise to the desired dynamics, and has the following form:
\begin{equation} \label{eq:RealisticAction}
S=S_{3BF}+S_{\text{grav}}+S_{\text{scal}}+S_{\text{Dirac}}+S_{\text{Yang-Mills}}+S_{\text{Higgs}}+S_{\text{Yukawa}}+S_{\text{spin}}+S_{\text{CC}}\,.
\end{equation}
Here we have:
\begin{eqnarray}
S_{3BF}&=&\int B_{\alpha}\wedge F^{\alpha}+ B^{[ab]}\wedge R_{[ab]}+e_a\wedge\nabla\beta^{a}+ \phi^{A}(\nabla\tilde{\gamma})_A+ \bar{\psi}_A(\nablar \gamma)^A-(\bar{\gamma} \nablal)_A\psi^A\,, \label{eq:3BFforStandardModel} \\
S_{\text{grav}}&=&-\int\lambda_{[ab]}\wedge\left(B^{[ab]}-\frac{1}{8\pi l_p^2}\varepsilon^{[ab]cd}e_c\wedge e_d\right)\,, \label{eq:gravConstraint} \\
S_{\text{scal}}&=&\int \tilde{\lambda}^{A}\wedge\left(\tilde{\gamma}_A-H_{abcA}e^a\wedge e^b\wedge e^c\right)+\Lambda^{abA}\wedge\left(H_{abcA}\varepsilon^{cdef}e_d\wedge e_e\wedge e_f-(\nabla\phi)_A\wedge e_a\wedge e_b\right)\,, \label{eq:scalConstraint} \\
S_{\text{Dirac}}&=&\int\bar{\lambda}_A\wedge\left(\gamma^A+\frac{i}{6}\varepsilon_{abcd}e^a\wedge e^b\wedge e^c\left(\gamma^d\psi\right)^{A}\right)-\lambda^{A}\wedge\left(\bar{\gamma}_A-\frac{i}{6}\varepsilon_{abcd}e^a\wedge e^b\wedge e^c \left(\bar{\psi}\gamma^d\right)_A\right)\,, \label{eq:DiracConstraint} \\
S_{\text{Yang-Mills}}&=&\int\lambda^{\alpha}\wedge\left(B_{\alpha}-12{C}_{\alpha\beta}M^{\beta}{}_{ab}e^a\wedge e^b\right)+\zeta_{\alpha}{}^{ab}\left(M^{\alpha}{}_{ab}\varepsilon_{cdef}e^c\wedge e^d \wedge e^e\wedge e^f-F^{\alpha}\wedge e_a\wedge e_b\right)\,, \label{eq:YangMillsConstraintTerm} \\
S_{\text{Higgs}}&=&-\int\frac{2}{4!}\chi\left(\phi^A\phi_A-v^2\right)^2\varepsilon_{abcd}e^a\wedge e^b\wedge e^c\wedge e^d\,, \label{eq:HiggsPotentialConstraint} \\
S_{\text{Yukawa}}&=&-\int \frac{2}{4!}Y_{ABC}\bar{\psi}^A\psi^B\phi^C\varepsilon_{abcd}e^a\wedge e^b\wedge e^c\wedge e^d\,, \label{eq:YukawaConstraint} \\
S_{\text{spin}}&=&\int 2\pi il_p^2\bar{\psi}_A\gamma_5\gamma^a\psi^A\varepsilon_{abcd}e^b\wedge e^c\wedge\beta^d\,, \label{eq:spinConstraint} \\
S_{\text{CC}}&=&-\int\frac{1}{96\pi l_p^2}\Lambda\varepsilon_{abcd}e^a\wedge e^b\wedge e^c\wedge e^d\,. \label{eq:CCconstraint}
\end{eqnarray}
In addition to the topological $3BF$ term (\ref{eq:3BFforStandardModel}), one can recognize:
\begin{itemize}
\item the gravitational constraint term (\ref{eq:gravConstraint}), giving rise to the gravitational degrees of freedom,
\item the scalar constraint (\ref{eq:scalConstraint}), giving rise to the massless scalar degrees of freedom,
\item the Dirac constriant (\ref{eq:DiracConstraint}), giving rise to the massless fermions,
\item the Yang-Mills constraint (\ref{eq:YangMillsConstraintTerm}), giving rise to the massless gauge bosons,
\item the Higgs potential constraint (\ref{eq:HiggsPotentialConstraint}), containing the self-interactions and the mass of the Higgs field,
\item the Yukawa constraint (\ref{eq:YukawaConstraint}), containing the interactions between the Higgs field and fermions, as well as fermion mixing matrices,
\item the spin constraint (\ref{eq:spinConstraint}), necessary for the appropriate coupling between fermion spins and torsion, and
\item the CC constraint (\ref{eq:CCconstraint}), introducing the cosmological constant.
\end{itemize}
The following free parameters are present in the action:
\begin{itemize}
\item $l_p$ is the Planck length, featuring in $S_\text{grav}$, $S_\text{spin}$ and $S_\text{CC}$,
\item $C_{\alpha\beta}$ represents the gauge coupling constant bilinear form, featuring in $S_\text{Yang-Mills}$,
\item $\chi$ is the coupling constant for the quartic self-interaction of the Higgs field, featuring in $S_\text{Higgs}$,
\item $v$ is the vacuum expectation value of the Higgs field, also featuring in $S_\text{Higgs}$,
\item $Y_{ABC}$ represent the Yukawa couplings and fermion mixing matrices, featuring in $S_\text{Yukawa}$, and
\item $\Lambda$ is the cosmological constant, featuring in $S_\text{CC}$.
\end{itemize}
The topological part $S_{3BF}$ and the constraints $S_\text{scal}$ and $S_\text{Dirac}$ do not contain any free parameters. Finally, note that the topological part $S_{3BF}$ is now rewritten in new notation. Specifically, $\cF$ is split into the internal symmetry field strength $F^\alpha$ (which is a function of the internal symmetry connection $\alpha^\alpha$) and the Riemann curvature two-form $R_{[ab]}$ (which is a function of the spin connection $\omega^{[ab]}$). The Lagrange multiplier $C$ is rewritten as the tetrad field one-form $e_a$, and the Lagrange multiplier $D$ is rewritten as a tuple of scalar and fermion fields $(\phi^A, \psi^A, \bar{\psi}_A)$. This change of notation also suggests the physical interpretation of the fields in (\ref{eq:3BFaction}).

Let us discuss the equations of motion for this action. After a certain amount of calculation, we obtain the equations solved for all Lagrange multiplier fields, in terms of the dynamical fields and their derivatives:
\begin{equation}
\begin{array}{rclcrcl}
    M_{\alpha ab}&=& \displaystyle -\frac{1}{48}\varepsilon_{abcd}F_{\alpha}{}^{\mu\nu}e^{c}{}_{\mu}e^d{}_{\nu}\,,&\hspace*{0.5cm}&
    \zeta^{\alpha ab}&=&\ds \frac{1}{4}{C}_{\beta}{}^{\alpha}\varepsilon^{abcd}F^{\beta}{}_{\mu\nu}e_c{}^{\mu}e_d{}^{\nu}\,, \vphantom{\ds\int} \\
    \lambda_{\alpha \mu\nu}&=&\ds -F_{\alpha \mu\nu}\,,&\hspace*{0.5cm}&
    B_{\alpha \mu\nu}&=&\ds -\frac{e}{2}{C}_{\alpha}{}^{\beta}\varepsilon_{\mu\nu\rho\sigma}F_{\beta}{}^{\rho\sigma}\,, \vphantom{\ds\int} \\
    \lambda_{[ab]\mu\nu}&=&\ds R_{[ab]\mu\nu}\,,&\hspace*{0.5cm}&
    B_{[ab]\mu\nu}&=&\ds \frac{1}{8\pi l_p^2}\varepsilon_{[ab]cd}e^c{}_{\mu}e^d{}_{\nu}\,, \vphantom{\ds\int}  \\
    \tilde{\lambda}^A{}_{\mu}&=&\ds \left(\nabla_{\mu}\phi\right)^A\,,&\hspace*{0.5cm}&
    \tilde{\gamma}^A{}_{\mu\nu\rho}&=&\ds -e\varepsilon_{\mu\nu\rho\sigma}\left(\nabla^{\sigma}\phi\right)^A\,, \vphantom{\ds\int} \\
    H^{abcA}&=&\ds \frac{1}{6e}\varepsilon^{\mu\nu\rho\sigma}\left(\nabla_{\mu}\phi\right)^Ae^a{}_{\nu}e^b{}_{\rho}e^c{}_{\sigma}\,,&\hspace*{0.5cm}&
    \Lambda^{abA}{}_{\mu}&=&\ds \frac{1}{6e}g_{\mu\lambda}\varepsilon^{\lambda \nu\rho\sigma}\left(\nabla_{\nu}\phi\right)^Ae^a{}_{\rho}e^b{}_{\sigma}\,, \vphantom{\ds\int} \\
    \gamma^A{}_{\mu\nu\rho}&=&\ds -i\varepsilon_{abcd}e^a{}_{\mu}e^b{}_{\nu}e^c{}_{\rho}\left(\gamma^d\psi\right)^A\,,&\hspace*{0.5cm}&
    \bar{\gamma}_{A\mu\nu\rho}&=&\ds i\varepsilon_{abcd}e^a{}_{\mu}e^b{}_{\nu}e^c{}_{\rho}\left(\bar{\psi}\gamma^d\right)_A\,, \vphantom{\ds\int} \\
    \lambda^A{}_{\mu}&=&\ds \left(\nablar_{\mu}\psi\right)^A\,,&\hspace*{0.5cm}&
    \bar{\lambda}_{A\mu}&=&\ds \left(\bar{\psi}\nablal_{\mu}\right)_A\,, \vphantom{\ds\int} \\
    \beta^a{}_{\mu\nu}&=&\ds 0\,.  \vphantom{\ds\int} 
    \end{array}
\end{equation}
Next we look at the equations of motion for the dynamical fields. The spin connection $\omega_{[ab]\mu}$ is not equivalent to the Levi-Civita connection, since fermionic fields give rise to nonzero torsion $T_a$: 
\begin{eqnarray}
    \omega_{[ab]\mu}&=&\Delta_{[ab]\mu}+K_{[ab]\mu}\,,\\ \label{spinskakoneksija}
    T_a \equiv \nabla e_a&=&2\pi i l_p^2 \, \varepsilon_{abcd} \, e^b\wedge e^c \, \bar{\psi}_A\gamma_5\gamma^d\psi^A=2\pi i l_p^2\, s_a\,.
\end{eqnarray}
Spin connection is represented as sum of Ricci rotation coefficients $\Delta_{[ab]\mu}$ and contorsion tensor $K_{[ab]\mu}$. Torsion 2-form is proportional to spin 2-form $s_a$, as usual in the Einstein-Cartan gravity.

The Einstein field equation has the usual form:
\begin{eqnarray}\label{ajn}
R_{\mu\nu}-\frac{1}{2}g_{\mu\nu}R+\Lambda g_{\mu\nu}=8\pi l_p^2 \, T_{\mu\nu}\,,
\end{eqnarray}
where the stress-energy tensor is given as:
\begin{eqnarray}
\nonumber
T_{\mu\nu}&=&F^{\alpha}{}_{\mu\rho}C_{\alpha}{}^{\beta}F_{\beta\nu}{}^{\rho}-\frac{1}{4}g_{\mu\nu}F^{\alpha}{}_{\rho\sigma}C_{\alpha}{}^{\beta}F_{\beta}{}^{\rho\sigma}\\
\nonumber
&+&\nabla_{\mu}\phi^A\nabla_{\nu}\phi_A-\frac{1}{2}g_{\mu\nu}\left(\nabla_{\rho}\phi^A\nabla^{\rho}\phi_A+2\chi\left(\phi^A\phi_A-v^2\right)^2\right)\\
&+&\frac{i}{2}\left(\bar{\psi}_A \nablalr_{\mu}\gamma_d\psi^A\right)e{}_{\nu}^d-\frac{1}{2}g_{\mu\nu}\left(i\left(\bar{\psi}_A \nablalr_{\rho}\gamma^d\psi^A\right)e_d{}^{\rho}-2Y_{ABC}\bar{\psi}^A\psi^B\phi^C\right)\,.
\end{eqnarray}
It features three parts, describing the Yang-Mills, scalar and fermion stress-energy, respectively.

Equations of motion for fermion and scalar fields are
\begin{eqnarray}
\left(i\gamma^{\mu} \nablar_{\mu}\delta_B^A-Y^A{}_{BC}\phi^C\right)\psi^B=0\,,\\
\bar{\psi}_B\left(\delta^B_A i \nablal_{\mu}\gamma^{\mu}+Y_{BAC}\phi^C\right)=0\,,\\
\nabla_{\mu}\nabla^{\mu}\phi^A-4\chi\left(\phi^B\phi_B-v^2\right)\phi^A=0\,,
\end{eqnarray}
while the equation of motion for Yang-Mills fields is:
\begin{equation}
\nabla_{\mu}F_{\alpha}{}^{\mu\nu}+\frac{1}{2}{C^{-1}}_{\alpha}{}^{\beta}\left(\triangleright_{\beta A B}\left(\phi^A\nabla^{\nu}\phi^B-\phi^B\nabla^{\nu}\phi^A\right)+i\bar{\psi}_A\psi_B\left(\triangleright_{\beta C}{}^A\gamma^{\nu CB}-\gamma^{\nu AC}\triangleright_{\beta C}{}^B\right)\right)=0\,.
\end{equation}
One can observe that all these equations of motion correspond precisely to the Standard Model coupled to Einstein-Cartan gravity, along with the cosmological constant term.

This completes the review of the realistic classical theory based on the constrained $3BF$ action and the $3$-group approach. In the next Section, we turn to the discussion of the symmetries of this model.

\section{\label{secIII}Constrained $3BF$ action and explicit symmetry breaking}

By adding simplicity constraints to the topological $3BF$ action, we reproduce the equations of motions for all known fields. But adding constraints also leads to breaking of the initial symmetry. In what follows, we will study each of the constraints separately, in order to determine which constraint breaks which symmetry group.  

The total symmetry group of the topological $3BF$ action has been studied in \cite{Radenkovic2022a,Djordjevic2023}, and has been shown to have the form $\cG_{3BF} = (\tilde{G}\ltimes(\tilde{H}_L\ltimes(\tilde{N}\times \tilde{M})))\ltimes HT$. The semidirect and direct products of groups $\tilde{G}$, $\tilde{H}_L$, $\tilde{M}$, $\tilde{N}$ correspond to the ordinary gauge symmetry of the action, while the $HT$ group corresponds to the so-called Henneaux-Teitelboim (HT) symmetry, which is trivial on-shell (for a review, see \cite{Djordjevic2023}).

The choice of the Standard Model $3$-group, introduced in the previous Section, implies however a more specific structure for the gauge group $\cG_{3BF}$. Namely, in the general case, the generators of the $\tilde{H}_L$ group can be naturally divided into the $\hat{H}$-generators and $\hat{L}$-generators, satisfying the Lie algebra commutation relations of the form
\begin{equation}
 [\hat{H}, \hat{H}] \sim \hat{L}\,, \qquad
 [\hat{H}, \hat{L}] \sim 0\,, \qquad
 [\hat{L}, \hat{L}] \sim 0\,,
\end{equation}
where the structure constants in the first commutator are proportional to the components of the Peiffer lifting map in the given $3$-group (see \cite{Radenkovic2022a} for a detailed analysis). Nevertheless, for our specific choice of the $3$-group the Peiffer lifting is trivial, implying that
\begin{equation}
 [\hat{H}, \hat{H}] \sim 0\,, \qquad
 [\hat{H}, \hat{L}] \sim 0\,, \qquad
 [\hat{L}, \hat{L}] \sim 0\,.
\end{equation}
This means that the group $\tilde{H}_L$ can be rewritten as a direct product
\begin{equation}
\tilde{H}_L = \tilde{H} \times \tilde{L}
\end{equation}
of two Abelian normal subgroups $\tilde{H}$ and $\tilde{L}$. Additionally, since in general the $\hat{H}$-generators are responsible for the semidirect product $\tilde{H}_L\ltimes(\tilde{N}\times \tilde{M})$ within $\cG_{3BF}$, with commutation relations of the form
\begin{equation}
 [\hat{H}, \hat{N}] \sim \hat{M}\,, \qquad
 [\hat{H}, \hat{M}] \sim 0\,, \qquad
 [\hat{L}, \hat{M}] \sim 0\,, \qquad
 [\hat{L}, \hat{N}] \sim 0\,,
\end{equation}
it is straightforward to conclude that in a case of any $3$-group with trivial Peiffer lifting, the symmetry group $\cG_{3BF}$ takes on a more specific form:
\begin{equation} \label{eq:SMgaugeGroup}
\cG_{3BF} = (\tilde{G}\ltimes( \tilde{L} \times (\tilde{H}\ltimes(\tilde{N}\times \tilde{M}))))\ltimes HT\,.
\end{equation}
Thus, (\ref{eq:SMgaugeGroup}) represents the gauge group of the $3BF$ theory based on the Standard Model $3$-group.

Let us now introduce the action of this group on the fields present in the $3BF$ action. For a general $3$-group, the infinitesimal transformations of the ordinary gauge part are derived in \cite{Radenkovic2022a} and listed as form-variations, while the infinitesimal transformations of the HT-part are defined in \cite{Djordjevic2023}. For the special case of the Standard Model $3$-group, the ordinary gauge transformations are given explicitly as follows:
\begin{equation} \label{eq:ordinaryGaugeTransf}
\begin{array}{rcl}
\delta_0^g\alpha^{\alpha}&=&- \nabla \epsilon_{\mathfrak{g}}{}^{\alpha}\,,\\
\delta_0^g\omega^{[ab]}&=&- \nabla \epsilon_{\mathfrak{g}}{}^{[ab]}\,,\\
\delta_0^g\beta^a&=&\triangleright_{\alpha b}{}^{a}\epsilon_{\mathfrak{g}}{}^{\alpha}\beta^b-\nabla\epsilon_{\mathfrak{h}}{}^{a}\,,\\
\delta_0^g\gamma^A&=&\triangleright_{\alpha B}{}^{A}\epsilon_{\mathfrak{g}}{}^{\alpha}\gamma^B+\nabla\epsilon_{\mathfrak{l}}{}^{A}\,,\\
\delta_0^g B^{\alpha}&=&f_{\beta\gamma}{}^{\alpha}\epsilon_{\mathfrak{g}}{}^{\beta}B^{\gamma}+e_a\wedge\epsilon_{\mathfrak{h}}{}^{b}\triangleright^{\alpha}\!{}_{b}{}^{a}-D_A\triangleright^\alpha\!\!{}_{B}{}^{A}\epsilon_{\mathfrak{l}}{}^{B}-\nabla\epsilon_{\mathfrak{m}}{}^{\alpha}+\beta_b\triangleright^{\alpha}\!{}_{a}{}^{b}\epsilon_{\mathfrak{n}}{}^{a}\,,\\
\delta_0^g B^{[ab]}&=&f_{[gh][ij]}{}^{[ab]}B^{[ij]}\epsilon_{\mathfrak{g}}{}^{[gh]}-\nabla\epsilon_{\mathfrak{m}}{}^{[ab]}
+e_c\wedge\epsilon_{\mathfrak{h}}{}^{d}\triangleright^{[ab]}{}_{d}{}^{c}+\beta_d\triangleright^{[ab]}{}_{c}{}^{d}\epsilon_{\mathfrak{n}}{}^{c}-\epsilon_{\mathfrak{l}}{}^{A}\triangleright^{[ab]}{}_{A}{}^{B}D_B\,,\\
\delta_0^g e^a&=&-\nabla\epsilon_{\mathfrak{n}}{}^{a}+\epsilon_{\mathfrak{g}}{}^{\alpha}\triangleright_{\alpha b}{}^{a}e^b\,,\\
\delta_0^g D^A&=&\triangleright_{\alpha B}{}^{A}\epsilon_{\mathfrak{g}}{}^{\alpha}D^B\,.
\end{array}
\end{equation}
The transformations are specified by the five free parameters, corresponding to their Lie algebras  --- the parameters $\epsilon_{\mathfrak{g}}{}^{\alpha}$ and $\epsilon_{\mathfrak{n}}{}^{a}$ are zero-forms, $\epsilon_{\mathfrak{h}}{}^{a}$ and $\epsilon_{\mathfrak{m}}{}^{\alpha}$ are one-forms, and $\epsilon_{\mathfrak{l}}{}^{A}$ are three-forms.

Regarding the HT symmetry, the infinitesimal transformations are most easily expressed in the following matrix form \cite{Djordjevic2023}:
{\footnotesize
\begin{equation} \label{eq:HTmatrix}
\left(
\begin{array}{lr}
\delta_{0}^{HT} B^{\alpha}{}_{\mu\nu} \vphantom{\ds\int} \\
\delta_{0}^{HT} C^{a}{}_{\mu} \vphantom{\ds\int} \\
\delta_{0}^{HT} D^{A} \vphantom{\ds\int} \\
\delta_{0}^{HT} \alpha^{\alpha}{}_{\mu} \vphantom{\ds\int} \\
\delta_{0}^{HT} \beta^{a}{}_{\mu\nu} \vphantom{\ds\int} \\
\delta_{0}^{HT} \gamma^{A}{}_{\mu\nu\rho} \vphantom{\ds\int} \\
\end{array}
\right) = \left(
\begin{array}{c|c|c|c|c|c}
\epsilon^{\alpha\beta}{}_{\mu\nu\sigma\lambda} & \epsilon^{\alpha b}{}_{\mu\nu\sigma} & \epsilon^{\alpha B}{}_{\mu\nu} &
\epsilon^{\alpha \beta}{}_{\mu\nu\sigma} & \epsilon^{\alpha b}{}_{\mu\nu\sigma\lambda} & \epsilon^{\alpha B}{}_{\mu\nu\sigma\lambda \xi} \vphantom{\ds\int} \\ \hline
\mu^{a \beta}{}_{\mu\sigma\lambda} & \epsilon^{a b}{}_{\mu\sigma} & \epsilon^{a B}{}_{\mu} &
\epsilon^{a \beta}{}_{\mu \sigma} & \epsilon^{a b}{}_{\mu \sigma\lambda} & \epsilon^{a B}{}_{\mu \sigma\lambda \xi} \vphantom{\ds\int} \\ \hline
\mu^{A \beta}{}_{\sigma\lambda} & \mu^{A b}{}_{\sigma} & \epsilon^{A B} &
\epsilon^{A \beta}{}_{\sigma} & \epsilon^{A b}{}_{\sigma\lambda} & \epsilon^{A B}{}_{\sigma\lambda \xi} \vphantom{\ds\int} \\ \hline
\mu^{\alpha \beta}{}_{\mu \sigma\lambda} & \mu^{\alpha b}{}_{\mu \sigma} & \mu^{\alpha B}{}_{\mu} &
\epsilon^{\alpha \beta}{}_{\mu \sigma} & \epsilon^{\alpha b}{}_{\mu \sigma\lambda} & \epsilon^{\alpha B}{}_{\mu \sigma\lambda \xi} \vphantom{\ds\int} \\ \hline
\mu^{a \beta}{}_{\mu \nu \sigma\lambda} & \mu^{a b}{}_{\mu \nu \sigma} & \mu^{a B}{}_{\mu \nu} &
\mu^{a \beta}{}_{\mu \nu \sigma} & \epsilon^{a b}{}_{\mu \nu \sigma\lambda} & \epsilon^{a B}{}_{\mu \nu \sigma\lambda \xi} \vphantom{\ds\int} \\ \hline
\mu^{A \beta}{}_{\mu \nu \rho \sigma\lambda} & \mu^{A b}{}_{\mu \nu \rho \sigma} & \mu^{A B}{}_{\mu \nu \rho} & 
\mu^{A \beta}{}_{\mu \nu \rho \sigma} & \mu^{A b}{}_{\mu \nu \rho \sigma\lambda} & \epsilon^{A B}{}_{\mu \nu \rho \sigma\lambda \xi} \vphantom{\ds\int} \\
\end{array}
\right) \left(
\begin{array}{c}
\frac{1}{2}\frac{\delta S}{\delta B^{\beta}{}_{\sigma \lambda}} \vphantom{\ds\int} \\
\frac{\delta S}{\delta C^{b}{}_{\sigma}} \vphantom{\ds\int} \\
\frac{\delta S}{\delta D^{B}} \vphantom{\ds\int} \\
\frac{\delta S}{\delta \alpha^{\beta}{}_{\sigma}} \vphantom{\ds\int} \\
\frac{1}{2}\frac{\delta S}{\delta \beta^{b}{}_{\sigma \lambda}} \vphantom{\ds\int} \\
\frac{1}{3!}\frac{\delta S}{\delta \gamma^{B}{}_{\sigma \lambda \xi}} \vphantom{\ds\int} \\
\end{array}
\right)\,.
\end{equation}
}%
Here, in order to ensure the antisymmetry of the parameter matrix, the following identities must hold:
\begin{linenomath}
\begin{equation} \label{eq:MuParamsFirstColumn}
\begin{array}{c}
\mu^{b \alpha}{}_{\sigma\mu\nu}=-\epsilon^{\alpha b}{}_{\mu\nu\sigma} \,, \qquad
\mu^{B \alpha}{}_{\mu\nu}=-\epsilon^{\alpha B}{}_{\mu\nu} \,, \qquad
\mu^{\beta\alpha}{}_{\sigma\mu\nu}=-\epsilon^{\alpha\beta}{}_{\mu\nu\sigma} \,, \vphantom{\ds\int} \\
\mu^{b \alpha}{}_{\sigma\lambda\mu\nu}=-\epsilon^{\alpha b}{}_{\mu\nu\sigma\lambda} \,, \qquad
\mu^{B \alpha}{}_{\sigma\lambda \xi \mu \nu}=-\epsilon^{\alpha B}{}_{\mu\nu\sigma\lambda\xi} \,. \vphantom{\ds\int} \\
\mu^{B a}{}_{\mu}=-\epsilon^{a B}{}_{\mu}\,, \qquad
\mu^{\beta a}{}_{\sigma\mu}=-\epsilon^{a \beta}{}_{\mu\sigma}\,, \vphantom{\ds\int} \\
\mu^{b a}{}_{\sigma\lambda\mu}=-\epsilon^{a b}{}_{\mu\sigma\lambda} \,, \qquad
\mu^{B a}{}_{\sigma\lambda\xi\mu}=-\epsilon^{a B}{}_{\mu\sigma\lambda\xi}\,, \vphantom{\ds\int} \\
\mu^{\beta A}{}_{\sigma}=-\epsilon^{A \beta}{}_{\sigma} \,, \qquad
\mu^{b A}{}_{\sigma\lambda}=-\epsilon^{A b}{}_{\sigma\lambda} \,, \qquad
\mu^{B A}{}_{\sigma\lambda\xi}=-\epsilon^{A B}{}_{\sigma\lambda\xi} \,, \vphantom{\ds\int} \\
\mu^{b \alpha}{}_{\sigma\lambda\mu}=-\epsilon^{\alpha b}{}_{\mu\sigma\lambda} \,, \qquad
\mu^{B \alpha}{}_{\sigma\lambda\xi\mu}=-\epsilon^{\alpha B}{}_{\mu\sigma\lambda\xi} \,, \qquad
\mu^{B a}{}_{\sigma\lambda\xi\mu\nu}=-\epsilon^{a B}{}_{\mu\nu\sigma\lambda\xi} \,.\end{array}
\end{equation}
\end{linenomath}
For more information about properties and importance of HT transformations, see \cite{Djordjevic2023}.

It is straightforward (if algebraically a bit involved) to verify that transformations (\ref{eq:ordinaryGaugeTransf}) and (\ref{eq:HTmatrix}) keep the topological action (\ref{eq:3BFaction}) invariant. However, this is not the case for the constrained action (\ref{eq:RealisticAction}), since each of the constraint terms may explicitly break one or more of these symmetries. In order to determine which symmetries are preserved and which are broken, and by which constraint term, we proceed as follows. For every individual constraint, we introduce the action
\begin{equation}
S = S_{3BF} + S_{\text{constraint}}\,,
\end{equation}
and we take the variation of this action with respect to (\ref{eq:ordinaryGaugeTransf}). The topological part $S_{3BF}$ is known to be already invariant, which means that the invariance of the action $S$ reduces to the requirement
\begin{equation} \label{eq:symmetryConstraintCondition}
\delta_0^g S_{\text{constraint}} = 0\,.
\end{equation}
This requirement may not be met automatically, but only by fixing the values of certain subset of parameters $\epsilon_{\mathfrak{g}}{}^{\alpha}$, $\epsilon_{\mathfrak{h}}{}^{a}$,  $\epsilon_{\mathfrak{l}}{}^{A}$, $\epsilon_{\mathfrak{m}}{}^{\alpha}$, and $\epsilon_{\mathfrak{n}}{}^{a}$. Each parameter that needs to be fixed indicates that the corresponding symmetry subgroup is broken by the constraint. In the following Subsections, we shall investigate each of the constraints (\ref{eq:gravConstraint})-(\ref{eq:CCconstraint}), and use (\ref{eq:symmetryConstraintCondition}) to discuss which symmetries are preserved and which are broken.

One should emphasize that the above method based on (\ref{eq:symmetryConstraintCondition}) makes sense only for the ordinary gauge symmetry, whereas the HT symmetry cannot be studied this way. Namely, as was explained in detail in \cite{Djordjevic2023}, the definition (\ref{eq:HTmatrix}) of the HT symmetry explicitly depends on the form of the action. This means that the very process of adding a constraint term to the action changes the HT symmetry group in a nontrivial way, most often by {\em increasing} its number of generators and parameters, so that the HT group of the constrained theory is {\em larger} than the HT group of the topological symmetry. This stands in sharp contrast to the ordinary gauge group, which is being broken down to one of its subgroups by the same process. Therefore, the question of explicit symmetry breaking by the introduction of the constraint term makes sense exclusively for the ordinary gauge symmetry, and cannot be even formulated for the HT symmetry.

\subsection{Gravitational simplicity constraint}
As explained above, the procedure for analyzing the symmetry breaking in the case of the gravitational simplicity constraint boils down to the calculation of the form variation of (\ref{eq:gravConstraint}) with respect to (\ref{eq:ordinaryGaugeTransf}) and then the discussion of the requirement (\ref{eq:symmetryConstraintCondition}). Specifically, we have:
\begin{equation}\label{eq:dsg}
\delta_0^g S_{\text{grav}}=-\int\left(\delta_0^g\lambda_{[ab]}\wedge\left(B^{[ab]}-\frac{1}{16\pi l_p^2}\varepsilon^{[ab]cd} e_c\wedge e_d\right)+\lambda_{[ab]}\wedge\left(\delta_0^gB^{[ab]}+\frac{2}{16\pi l_p^2}\varepsilon^{[ab]cd}\delta_0^g e_c\wedge e_d \right)\right)\,.
\end{equation}
The variation of the Lagrange multiplier $\lambda_{[ab]}$ is not defined initially, so we choose to define it in such way to preserve as many symmetries as we can. Substituting (\ref{eq:ordinaryGaugeTransf}) into (\ref{eq:dsg}), after some algebra, the variation of the gravitational constraint becomes:
\begin{eqnarray}\label{eq:sag}
\nonumber
\delta_0^g S_{\text{grav}}&=&\int\left(\delta^g_0\lambda_{[ij]}-\lambda_{[i|h}\epsilon_{\mathfrak{g}}{}_{|j]}{}^{h}\right)\wedge \left(B^{[ij]}-\frac{1}{16\pi l_p^2}\varepsilon^{[ij]nm}e_n\wedge e_m\right)\\
&+&\lambda_{[ab]}\wedge e_d\wedge\left(\epsilon_{\mathfrak{h}}{}^{[a|}\eta^{|b]d}-\frac{1}{8\pi l_p^2}\varepsilon^{[ab]cd}\eta_{fc}\nabla\epsilon_{\mathfrak{n}}{}^{f}\right)+\lambda_{[ab]}\wedge\left(\epsilon_{\mathfrak{n}}{}^{[a|}\beta^{|b]}-\nabla\epsilon_{\mathfrak{m}}{}^{[ab]}-\epsilon_{\mathfrak{l}}{}^{A}\triangleright^{[ab]}{}_{A}{}^{B}D_B\right)\,,
\end{eqnarray}
from where we can see that we can choose the variation of the multiplier $\lambda_{[ab]}$ as follows:
\begin{equation}\label{eq:glambda}
\delta_0^g\lambda_{[ij]}=-\lambda_{[ab]}f_{[gh][ij]}{}^{[ab]}\epsilon_{\mathfrak{g}}{}^{[gh]}=\lambda_{[i|h}\epsilon_{\mathfrak{g}}{}_{|j]}{}^{h}\,.
\end{equation}
This choice removes the whole first row in (\ref{eq:sag}). However, from the second row one can see that the requirement (\ref{eq:symmetryConstraintCondition}) can only be satisfied if one chooses specific values of $\epsilon_{\mathfrak{h}}{}^{a}$, $\epsilon_{\mathfrak{n}}{}^{a}$, $\epsilon_{\mathfrak{m}}{}^{\alpha}$ and $\epsilon_{\mathfrak{l}}{}^{A}$. The only parameter that is not fixed is $\epsilon_{\mathfrak{g}}{}^{\alpha}$. This implies that this constraint breaks all symmetry groups $\tilde{M}$, $\tilde{N}$, $\tilde{L}$, and $\tilde{H}$, except for the group $\tilde{G}$, which remains unbroken.


\subsection{Scalar simplicity constraint}
Using the above procedure we examine all the remaining constraints. In the case of the constraint for the scalar field, we have:
\begin{eqnarray}\label{skalarno}
\nonumber
\delta_0^gS_{\text{scal}}&=&\int\Big[\delta_0^g\tilde{\lambda}_A\wedge\left(\tilde{\gamma}^A-H_{abc}{}^A e^a\wedge e^b\wedge e^c\right) \vphantom{\ds\int} \\
\nonumber
&+&\tilde{\lambda}_A\wedge\left(\delta_0^g\tilde{\gamma}^A-\delta_0^gH_{abc}{}^A e^a\wedge e^b\wedge e^c-3H_{abc}{}^A\delta_0^ge^a\wedge e^b \wedge e^c\right)\vphantom{\ds\int}\\
&+&\delta_0^g\Lambda^{ab}{}_A\wedge\left(H_{abc}{}^A\varepsilon^{cdef}e_d\wedge e_e\wedge e_f-\nabla\phi^A\wedge e_a\wedge e_b\right)\vphantom{\ds\int}\\
\nonumber
&+&\Lambda^{ab}{}_A\wedge\Big(\delta_0^gH_{abc}{}^A\varepsilon^{cdef}e_d\wedge e_e\wedge e_f+3H_{abc}{}^A\varepsilon^{cdef}\delta_0^ge_d\wedge e_e\wedge e_f\vphantom{\ds\int}\\
\nonumber
&-&\nabla\delta_0^g\phi^A\wedge e_a\wedge e_b-\delta_0^g\alpha^{[kl]}\triangleright_{[kl]B}{}^{A}\phi^B\wedge e_a\wedge e_b-2\nabla\phi^A\wedge\delta_0^g e_a\wedge e_b\Big)\Big]\,.\vphantom{\ds\int}
\end{eqnarray}
Substituting (\ref{eq:ordinaryGaugeTransf}) into (\ref{skalarno}), we get:
\begin{eqnarray} \label{eq:skalarnodva}
\nonumber
\delta_0^gS_{\text{scal}}&=&\int\Big[(\delta_0^g\tilde{\lambda}_A+\tilde{\lambda}_B\epsilon_{\mathfrak{g}}{}^{[ij]}\triangleright_{[ij]A}{}^{B})\wedge\left(\tilde{\gamma}^A-H_{abc}{}^A e^a\wedge e^b\wedge e^c\right)\vphantom{\ds\int}\\
\nonumber
&-&\left(\delta_0^g H_{abc}{}^A-\epsilon_{\mathfrak{g}}{}^{[ij]} H_{abc}{}^B\triangleright_{[ij]B}{}^{A}\right)(\tilde{\lambda}_A\eta^{ad}\eta^{be}\eta^{cf}-\Lambda^{ab}{}_A\varepsilon^{cdef})\wedge e_d\wedge e_e\wedge e_f\vphantom{\ds\int}\\
&+&(\delta_0^g\Lambda^{ab}{}_A+\Lambda^{ab}{}_B\epsilon_{\mathfrak{g}}{}^{[cd]}\triangleright_{[cd]A}{}^{B})\wedge\left(H_{abc}{}^A\varepsilon^{cdef}e_d\wedge e_e\wedge e_f-\nabla\phi^A\wedge e_a\wedge e_b\right)\vphantom{\ds\int}\\
\nonumber
&+&(\tilde{\lambda}_A\eta^{ad}\eta^{be}\eta^{cf}-\Lambda^{ab}{}_A\varepsilon^{cdef})\wedge 3H_{abc}{}^A(\nabla\epsilon_{\mathfrak{n}d})\wedge e_e\wedge e_f\vphantom{\ds\int}\\
\nonumber
&+&2\Lambda^{ab}{}_A\wedge\nabla\phi^A\wedge\nabla\epsilon_{\mathfrak{n}a}\wedge e_b+\tilde{\lambda}_A\wedge\nabla\epsilon_{\mathfrak{l}}{}^{A}\Big]\vphantom{\ds\int}\,.
\end{eqnarray}
It is obvious from fourth and fifth row from above equation that scalar field constraint breaks only $\tilde{N}$ and $\tilde{L}$ symmetries, while $\tilde{H}$ and $\tilde{M}$ symmetries are preserved since their parameters $\epsilon_{\mathfrak{h}}{}^{a}$ and $\epsilon_{\mathfrak{m}}{}^{\alpha}$ do not even appear in the variation (\ref{eq:skalarnodva}). Finally, in order to preserve $\tilde{G}$ symmetry, we choose to define variations of new multipliers as:
\begin{equation}
\delta_0^g\tilde{\lambda}^A=\epsilon_{\mathfrak{g}}{}^{[ij]}\tilde{\lambda}^B\triangleright_{[ij]B}{}^{A}\,,\qquad
\delta_0^gH_{abc}{}^A=H_{abc}{}^B \epsilon_{\mathfrak{g}}{}^{[ij]} \triangleright_{[ij]B}{}^{A}\,,\qquad
\delta_0^g\Lambda^{abA}=\epsilon_{\mathfrak{g}}{}^{[cd]}\Lambda^{abB}\triangleright_{[cd]B}{}^{A}\,.
\end{equation}


\subsection{Dirac simplicity constraint}

In the same way, variation of the constraint for the Dirac fields gives us:
\begin{eqnarray}\label{dir}
\nonumber
\delta_0^gS_{\text{Dirac}}&=&\int(\delta_0^g\bar{\lambda}_{A})\wedge\left(\gamma^{A}+\frac{i}{6}\varepsilon_{abcd}e^a\wedge e^b\wedge e^c\left(\gamma^d\psi\right)^{A}\right)\\
\nonumber
&-&(\delta_0^g\lambda^{A})\wedge\left(\bar{\gamma}_{A}-\frac{i}{6}\varepsilon_{abcd}e^a\wedge e^b\wedge e^c\left(\bar{\psi}\gamma^d\right)_{A}\right)\\
\nonumber
&+&\bar{\lambda}_{A}\wedge\left((\delta_0^g\gamma^{A})+\frac{i}{6}\varepsilon_{abcd}e^a\wedge e^b\wedge e^c\left(\gamma^d(\delta_0^g\psi)\right)^{A}+\frac{i}{2}\varepsilon_{abcd}(\delta_0^g e^a)\wedge e^b\wedge e^c\left(\gamma^d\psi\right)^{A}\right)\\
&-&\lambda^{A}\wedge\left((\delta_0^g\bar{\gamma}_{A})-\frac{i}{6}\varepsilon_{abcd}e^a\wedge e^b\wedge e^c\left(\gamma^d(\delta_0^g\bar{\psi})\right)_{A}-\frac{i}{2}\varepsilon_{abcd}(\delta_0^g e^a)\wedge e^b\wedge e^c\left(\gamma^d\bar{\psi}\right)_{A}\right)\,.
\end{eqnarray}
Substituting (\ref{eq:ordinaryGaugeTransf}) into (\ref{dir}) gives us:
\begin{eqnarray}
\nonumber
\delta_0^gS_{\text{Dirac}}&=&\int(\delta_0^g\bar{\lambda}_{A}+\epsilon_{\mathfrak{g}}{}^{\alpha}\triangleright_{\alpha A}{}^{B}\bar{\lambda}_{B})\wedge\left(\gamma^{A}+\frac{i}{6}\varepsilon_{abcd}e^a\wedge e^b\wedge e^c\left(\gamma^d\psi\right)^{A}\right)\\
\nonumber
&-&(\delta_0^g\lambda^{A}+\epsilon_{\mathfrak{g}}{}^{\alpha}\triangleright_{\alpha}{}^{A}{}_{B}\lambda^{B})\wedge\left(\bar{\gamma}_{A}-\frac{i}{6}\varepsilon_{abcd}e^a\wedge e^b\wedge e^c\left(\bar{\psi}\gamma^d\right)_{A}\right)\\
\nonumber
&+&\bar{\lambda}_{A}\wedge\left(\nabla\epsilon_{\mathfrak{l}}{}^{A}+\frac{i}{2}\varepsilon_{abcd}(\nabla\epsilon_{\mathfrak{n}}{}^{a})\wedge e^b\wedge e^c\left(\gamma^d\psi\right)^{A}\right)\\
&-&\lambda^{A}\wedge\left(\nabla\bar{\epsilon}_{\mathfrak{l}A}-\frac{i}{2}\varepsilon_{abcd}(\nabla\epsilon_{\mathfrak{n}}{}^{a})\wedge e^b\wedge e^c\left(\bar{\psi}\gamma^d\right)_{A}\right)\,.
\end{eqnarray}
This constraint also breaks only $\tilde{N}$ and $\tilde{L}$ symmetries, similar to the scalar constraint. The variation laws for new multipliers are chosen to be: 
\begin{equation}
\delta_0^g\bar{\lambda}_{A}=\epsilon_{\mathfrak{g}}{}^{\alpha}\triangleright_{\alpha}{}^{B}{}_{A}\bar{\lambda}_{B}\,,\qquad
\delta_0^g\lambda^{A}=\epsilon_{\mathfrak{g}}{}^{\alpha}\triangleright_{\alpha}{}^{A}{}_{B}\lambda^{B}\,.
\end{equation}

\subsection{Yang-Mills simplicity constraint}

Yang-Mills simplicity constraint is similar to gravitational constraint, but it contains two more Lagrange multipliers. We apply the same variation procedure as above:
\begin{eqnarray}\label{yan}
\nonumber
\delta_0^g S_{\text{Yang-Mills}}&=&\int \delta_0^g\lambda^{\alpha}\wedge\left(B_{\alpha}-12{C}^{\alpha\beta}M_{\beta ab}e^a\wedge e^b\right)\vphantom{\ds\int}\\
\nonumber
&+&\lambda^{\alpha}\wedge\left(\delta_0^g B_{\alpha}-12{C}^{\alpha\beta}\delta_0^g M_{\beta ab}e^a\wedge e^b-24{C}^{\alpha\beta}M_{\beta ab}\delta_0^g e^a\wedge e^b\right)\vphantom{\ds\int}\\
&+&\delta_0^g\zeta^{\alpha ab}\left(M_{\alpha ab}\varepsilon_{cdef}e^c\wedge e^d\wedge e^e\wedge e^f-F_{\alpha}\wedge e_a\wedge e_b\right)\vphantom{\ds\int}\\
\nonumber
&+&\zeta^{\alpha ab}\left(\left(\delta_0^g M_{\alpha ab}\right)\varepsilon_{cdef}e^c\wedge e^d \wedge e^e\wedge e^f+4M_{\alpha ab}\varepsilon_{cdef}\left(\delta_0^g e^c\right)\wedge e^d\wedge e^e\wedge e^f\right.\vphantom{\ds\int}\\
\nonumber
&-&\left.\left(\delta_0^g F_{\alpha}\right)\wedge e_a\wedge e_b-2F_{\alpha}\wedge \left(\delta_0^g e_a\right)\wedge e_b\right)\vphantom{\ds\int}\,,
\end{eqnarray}
where
\begin{equation}\label{yanpaket}
\delta_0^g F_{\alpha}=\epsilon_{\mathfrak{g}}{}^{\beta}F^{\gamma}\triangleright_{\alpha\beta\gamma}\vphantom{\ds\int}\,.
\end{equation}
The variation of the field strength (\ref{yanpaket}) is obtained by varying the definition (\ref{eq:ThreeCurvatureDef}) using (\ref{eq:ordinaryGaugeTransf}).
Combining equations (\ref{eq:ordinaryGaugeTransf}), (\ref{yan}), and (\ref{yanpaket}) gives us:
\begin{eqnarray}
\nonumber
\delta_0^g S_{\text{Yang-Mills}}&=&\int \left(\delta_0^g\lambda^{\alpha}+\lambda^{\gamma}\epsilon_{\mathfrak{g}}{}^{\beta}\triangleright_{\gamma\beta}{}^{\alpha}\right)\wedge\left(B_{\alpha}-12{C}^{\alpha\beta}M_{\beta ab}e^a\wedge e^b\right)\vphantom{\ds\int}\\
\nonumber
&+&\left(\delta_0^g M_{\alpha ab}-\epsilon_{\mathfrak{g}}{}^{\beta}M^{\gamma}{}_{ab}\triangleright_{\alpha\beta\gamma}\right)\left(\zeta^{\alpha ab}\varepsilon_{cdef} e^c\wedge e^d\wedge e^e\wedge e^f-12{C}^{\alpha}{}_{\beta}\lambda^{\beta}\wedge e^a\wedge e^b\right)\vphantom{\ds\int}\\
&+&\left(\delta_0^g\zeta^{\alpha ab}+\triangleright_{\gamma\beta}{}^{\alpha}\zeta^{\gamma ab}\epsilon_{\mathfrak{g}}{}^{\beta}\right)\left(M_{\alpha ab}\varepsilon_{cdef}e^c\wedge e^d\wedge e^e\wedge e^f-F_{\alpha}\wedge e_a\wedge e_b\right)\vphantom{\ds\int}\\
\nonumber
&+&\lambda^{\alpha}\wedge\left(-\nabla\epsilon_{\mathfrak{m}\alpha}-\epsilon_{\mathfrak{l}}{}^{A}\triangleright_{\alpha A}{}^{B}D_B+\epsilon_{\mathfrak{n}}{}^{a}\triangleright_{\alpha a}{}^{b}\beta_b-\epsilon_{\mathfrak{h}}{}^{a}\wedge e_b\triangleright_{\alpha a}{}^{b}+24{C}^{\alpha\beta}M_{\beta ab}\left(\nabla\epsilon_{\mathfrak{n}}{}^{a}\right)\wedge e^b\right)\vphantom{\ds\int}\\
\nonumber
&+&\zeta^{\alpha ab}\left(-4M_{\alpha ab}\varepsilon_{cdef}\left(\nabla\epsilon_{\mathfrak{n}}{}^{c}\right)\wedge e^d\wedge e^e\wedge e^f+2F_{\alpha}\wedge (\nabla\epsilon_{\mathfrak{n}a})\wedge e_b\right)\,.\vphantom{\ds\int}
\end{eqnarray}
Similar to the case of gravitational constraint, all symmetries, $\tilde{H}$, $\tilde{L}$, $\tilde{N}$ and $\tilde{M}$, except for $\tilde{G}$, are broken. The variations for new multipliers are chosen as:
\begin{equation}
\delta_0^g\lambda^{\alpha}=\epsilon_{\mathfrak{g}}{}^{\beta}\lambda^{\gamma}\triangleright_{\beta\gamma}{}^{\alpha}\,,\qquad
\delta_0^g M_{\alpha ab}=\epsilon_{\mathfrak{g}}{}^{\beta}M^{\gamma}{}_{ab}\triangleright_{\alpha\beta\gamma}\,,\qquad
\delta_0^g\zeta^{\alpha ab}=\epsilon_{\mathfrak{g}}{}^{\beta}\zeta^{\gamma ab}\triangleright_{\beta\gamma}{}^{\alpha}\,.
\end{equation}

\subsection{Higgs, Yukawa, spin, and CC terms}

Variation of Higgs term gives us
\begin{eqnarray}
\nonumber
\delta_0^g S_{\text{Higgs}}&=&-\frac{1}{3}\chi\int2(\phi_A\phi^A-v^2)\phi_A(\delta_0^{g}\phi^A)\varepsilon^{abcd}e_a\wedge e_b\wedge e_c\wedge e_d\\
&+&(\phi_A\phi^A-v^2)^2\varepsilon^{abcd}(\delta_0^{g}e_a)\wedge e_b\wedge e_c\wedge e_d\,.\vphantom{\ds\int}
\end{eqnarray}
This implies that
\begin{equation} \label{eq:HiggsConstraintVariation}
\delta_0^g S_{\text{Higgs}}=\frac{1}{3}\chi\int(\phi_A\phi^A-v^2)^2\varepsilon^{abcd}(\nabla\epsilon_{\mathfrak{n}a})\wedge e_b\wedge e_c\wedge e_d\,,
\end{equation}
where we have used the identity $\triangleright_{\alpha AB}=-\triangleright_{\alpha BA}$ from the Theorem. This constraint does not break $\tilde{G}$ symmetry since its parameter drops out of (\ref{eq:HiggsConstraintVariation}) despite being present in the form variations for both $\phi^A$ and $e_a$, so only $\tilde{N}$ symmetry is broken.

The variation of Yukawa coupling term is:
\begin{eqnarray}\label{ykava}
\nonumber
\delta_0^g S_{\text{Yukawa}}&=&-\frac{2}{4!}\int Y_{ABC}\delta_0^g(\bar{\psi}^A\psi^B)\phi^C\varepsilon_{abcd}e^a\wedge e^b \wedge e^c\wedge e^d+Y_{ABC}\bar{\psi}^A\psi^B\delta_0^g\phi^C\varepsilon_{abcd}e^a\wedge e^b \wedge e^c\wedge e^d\vphantom{\ds\int}\\
&+&4Y_{ABC}\bar{\psi}^A\psi^B\phi^C\varepsilon_{abcd}\delta_0^g e^a\wedge e^b \wedge e^c\wedge e^d\,.\vphantom{\ds\int}
\end{eqnarray}
Again, substituting the variations of fields (\ref{eq:ordinaryGaugeTransf}) into (\ref{ykava}) gives:
\begin{eqnarray}
\delta_0^g S_{\text{Yukawa}}&=&\frac{1}{3}\int Y_{ABC}\bar{\psi}^A\psi^B\phi^C\varepsilon_{abcd}(\nabla\epsilon_{\mathfrak{n}}{}^{a})\wedge e^b \wedge e^c\wedge e^d\,.\vphantom{\ds\int}
\end{eqnarray}
As above, only $\tilde{N}$ symmetry is broken, because
the $Y_{ABC}$ matrix is defined in such a way to preserve $\tilde{G}$ symmetry.

Spin coupling term does not break $\tilde{G}$ symmetry for the same reasons as the Dirac or Higgs terms, but it does break $\tilde{N}$ and $\tilde{H}$ symmetries:
\begin{eqnarray}
\nonumber
\delta_0^g S_{\text{spin}}&=&2\pi i l_p^2\varepsilon_{abcd}\int\left(\delta_0^g(\bar{\psi}\gamma_5\gamma^a\psi)e^b\wedge e^c\wedge\beta^d+\bar{\psi}\gamma_5\gamma^a\psi\delta_0^g(e^b\wedge e^c\wedge\beta^d)\right)\vphantom{\ds\int}\,,
\\
&=&-2\pi i l_p^2\varepsilon_{abcd}\int\bar{\psi}\gamma_5\gamma^a\psi\left(2(\nabla\epsilon_{\mathfrak{n}}{}^{b})\wedge e^c\wedge\beta^d+e^b\wedge e^c\wedge(\nabla\epsilon_{\mathfrak{h}}{}^{d})\right)\vphantom{\ds\int}\,.
\end{eqnarray}

Finally, the CC term breaks only $\tilde{N}$ symmetry:
\begin{eqnarray}
\delta_0^g S_{\text{CC}}=-\int\frac{1}{24\pi l_p^2}\Lambda\varepsilon_{abcd}\delta_0^g e^a\wedge e^b\wedge e^c\wedge e^d=-\int\frac{1}{24\pi l_p^2}\Lambda\varepsilon_{abcd}\nabla \epsilon_{\mathfrak{n}}{}^{a}\wedge e^b\wedge e^c\wedge e^d\vphantom{\ds\int}\,.
\end{eqnarray}

\subsection{Overview of symmetry breaking}\label{endofens}

Summing up all the results from this Section, we can make a table of symmetries and constraints. Each field labeled with $\times$ corresponds to the breaking of a given symmetry by a given constraint term:

\begin{center}
\begin{tabular}{| c || c | c | c | c | c | c | c | c |}
\hline
\, & $S_{\text{grav}}$ & $S_{\text{scal}}$ & $S_{\text{Dirac}}$ & $S_{\text{Yang-Mills}}$ & $S_{\text{Higgs}}$ & $S_{\text{Yukawa}}$ & $S_{\text{spin}}$ & $S_{\text{CC}}$ \\
\hline\hline
$\tilde{G}$ & \, & \, & \, & \, & \, & \, & \, & \, \\
\hline
$\tilde{H}$ & $\times$ & $\,$ & $\,$ & $\times $ & & \, & $\times$ & \, \\
\hline
$\tilde{L}$ & $\times$ & $\times$ & $\times$ & $\times$ & \, & \, & \, & \, \\
\hline
$\tilde{M}$ & $\times$ & \, & \, & $\times$ & \, & \, & \, & \, \\
\hline
$\tilde{N}$ & $\times$ & $\times$ & $\times$ & $\times$ & $\times$ & $\times$ & $\times$ & $\times$ \\
\hline
\end{tabular}
\end{center}
From the above table one can observe several interesting features. First, $\tilde{G}$ symmetry is preserved by all constraints, while $\tilde{N}$ symmetry is broken by all constraints. Second, the gravitational and Yang-Mills constraints break all symmetries except $\tilde{G}$, and these constraints are the only ones to do so. Finally, the Higgs potential constraint, the Yukawa coupling constraint and the cosmological constant constraint break exclusively the $\tilde{N}$ symmetry, while preserving all others.

In addition to the above results, there are three more constraint terms, which do not appear in the action (\ref{eq:RealisticAction}) but will appear later on in Sections \ref{secIV} and \ref{secV}, after we rewrite the action in a form corresponding to the spontaneously broken symmetry. The first of these is the mass term for scalar fields:
\begin{equation}\label{skalarna_masa}
S_{\text{scalar mass}}=-\frac{m^2}{4!}\varepsilon_{abcd}\int\phi_A\phi^Ae^a\wedge e^b\wedge e^c\wedge e^d\,.
\end{equation}
Variation of this term is
\begin{eqnarray}
\delta_0^g S_{\text{scalar mass}}&=&-\frac{m^2}{4!}\varepsilon_{abcd}\int\left(2(\delta_0^g\phi_A)\phi^Ae^a\wedge e^b\wedge e^c\wedge e^d+4\phi_A\phi^A(\delta_0^ge^a)\wedge e^b\wedge e^c\wedge e^d\right)\,,\vphantom{\ds\int}
\end{eqnarray}
which reduces to:
\begin{eqnarray}
&&\delta_0^g S_{\text{scalar mass}}=\frac{m^2}{3!}\varepsilon_{abcd}\int\phi_A\phi^A(\nabla\epsilon_{\mathfrak{n}}{}^{a})\wedge e^b\wedge e^c\wedge e^d\,.\vphantom{\ds\int}
\end{eqnarray}
We conclude that the term (\ref{skalarna_masa}) breaks only $\tilde{N}$ symmetry.

The second new constraint term is the Dirac mass term:
\begin{equation}\label{Dirakova masa}
S_{\text{Dirac mass}}=-\frac{m}{12}\varepsilon_{abcd}\int\bar{\psi}_A\psi^Ae^a\wedge e^b\wedge e^c\wedge e^d\,.
\end{equation}
Its variation is
\begin{eqnarray}
\nonumber
\delta_0^g S_{\text{Dirac mass}}&=&-\frac{m}{12}\varepsilon_{abcd}\int\left((\delta_0^g\bar{\psi}_A)\psi^Ae^a\wedge e^b\wedge e^c\wedge e^d+\bar{\psi}_A(\delta_0^g\psi^A)e^a\wedge e^b\wedge e^c\wedge e^d\right.\vphantom{\ds\int}\\
&+&\left.4\bar{\psi}_A\psi^A(\delta_0^ge^a)\wedge e^b\wedge e^c\wedge e^d\right)\,,\vphantom{\ds\int}
\end{eqnarray}
which reduces to:
\begin{eqnarray}
&&\delta_0^g S_{\text{Dirac mass}}=\frac{m}{3}\varepsilon_{abcd}\int\bar{\psi}_A\psi^A(\nabla\epsilon_{\mathfrak{n}}{}^{a})\wedge e^b\wedge e^c\wedge e^d\,.\vphantom{\ds\int}
\end{eqnarray}
Similar to the scalar mass term, this constraint also breaks only $\tilde{N}$ symmetry.

The third new constraint term is the Proca constraint. This term will explicitly appear in Section \ref{secIV} in equation (\ref{eq:ProcaConstraintTerm}) below, as part of the discussion of the Proca action. It has the following form (see Section \ref{secIV} for the details of the notation):
\begin{equation}
S_{\text{Proca}} = \int \Theta^{\alpha ab}\wedge \left(\Xi_{\alpha abc}\varepsilon^{cdef}e_d\wedge e_e \wedge e_f+\frac{M}{g} \alpha_\alpha\wedge e_a \wedge e_b\right) + \frac{M}{g}\alpha^{\alpha}\wedge \Xi_{\alpha abc}e^a\wedge e^b\wedge e^c \,.
\end{equation}
The variation of this term is:
\begin{eqnarray}
\nonumber
\delta_0^g S_{\text{Proca}}&=&\int\Bigg[\frac{M}{g}\left(\delta_0^g\alpha^{\alpha}\wedge \Xi_{\alpha abc}e^a+\alpha^{\alpha}\wedge\delta_0^g\Xi_{\alpha abc}e^a+3\alpha^{\alpha}\wedge \Xi_{\alpha abc}\delta_0^g e^a\right)\wedge e^b\wedge e^c\vphantom{\ds\int}\\
\nonumber
&+&\delta_0^g\Theta^{\alpha ab}\wedge\left(\Xi_{\alpha abc}\varepsilon^{cdef}e_d\wedge e_e\wedge e_f+\frac{M}{g}\alpha_{\alpha}\wedge e_a\wedge e_b\right)\vphantom{\ds\int}\\
\nonumber
&+&\Theta^{\alpha ab}\wedge\left(\delta_0^g\Xi_{\alpha abc}\varepsilon^{cdef}e_d\wedge e_d\wedge e_e\wedge e_f+3\Xi_{\alpha abc}\varepsilon^{cdef}\delta_0^g e_d\wedge e_e\wedge e_f\right)\vphantom{\ds\int}\\
&+&\frac{M}{g}\Theta^{\alpha ab}\wedge\left(\delta_0^g\alpha_{\alpha}\wedge e_a\wedge e_b+2\alpha_{\alpha}\wedge \delta_0^g e_a\wedge e_b\right)\vphantom{\ds\int}\Bigg]\,.
\end{eqnarray}
Substituting the variations of connection $\alpha$ and tetrad fields, and using the fact that $\triangleright_{\alpha a}{}^b=0$, we get:
\begin{eqnarray}
\nonumber
\delta_0^g S_{\text{Proca}}&=&\int\Bigg[\delta_0^g\Theta^{\alpha ab}\wedge\left(\Xi_{\alpha abc}\varepsilon^{cdef}e_d\wedge e_e\wedge e_f+\frac{M}{g}\alpha_\alpha\wedge e_a\wedge e_b\right)\\
\nonumber
&+&\delta_0^g\,\Xi_{\alpha abc}\left(\frac{M}{g}\alpha^\alpha\wedge e^a\wedge e^b\wedge e^c+\Theta^{\alpha ab}\wedge\varepsilon^{cdef}e_d\wedge e_e\wedge e_f\right)\\
\nonumber
&+&\frac{M}{g}\nabla\epsilon_{\mathfrak{g}}{}^{\alpha}\wedge\left(\Theta^{\alpha ab}\wedge e_a\wedge e_b-\Xi_{\alpha abc}e^a\wedge e^b\wedge e^c\right)\\
\nonumber
&-&3\,\Xi_{\alpha abc}\left(\frac{M}{g}\alpha^{\alpha}\wedge\nabla\epsilon_{\mathfrak{n}}{}^{a}\wedge e^b\wedge e^c+\Theta^{\alpha ab}\varepsilon^{cdef}\nabla\epsilon_{\mathfrak{n}}{}_{d}\wedge e_e\wedge e_f\right)\\
&-&2\frac{M}{g}\Theta^{\alpha ab}\wedge\alpha_\alpha\wedge\nabla\epsilon_{\mathfrak{n}}{}_{a}\wedge e_b\vphantom{\ds\int}\Bigg]\,.
\end{eqnarray}
We can eliminate the first two rows by choosing the variations of the new multipliers to be
\begin{equation}
\delta_0^g\Theta^{\alpha ab}=0 \,, \qquad \delta_0^g\,\Xi_{\alpha abc}=0\,.
\end{equation} 
However, in addition to broken $\tilde{N}$ symmetry, the Proca constraint is the only constraint which breaks $\tilde{G}$ symmetry, since its parameter appears explicitly in the third row.

Finally, let us note that the scalar mass constraint, Dirac mass constraint and the Proca constraint supplement the above table of constraints with three more columns, as follows:
\begin{center}
\begin{tabular}{| c || c | c | c |}
\hline
\, & $S_{\text{scalar mass}}$ & $S_{\text{Dirac mass}}$ & $S_{\text{Proca}}$ \\
\hline\hline
$\tilde{G}$& \, & \, & $\times$ \\
\hline
$\tilde{H}$ & \, & \, & \, \\
\hline
$\tilde{L}$ & \, & \, & \, \\
\hline
$\tilde{M}$ & \, & \, & \, \\
\hline
$\tilde{N}$ & $\times$ & $\times$ & $\times$ \\
\hline
\end{tabular}
\end{center}

This concludes the analysis of explicit symmetry breaking of the constrained $3BF$ theory. In what follows, we turn to the detailed analysis of the Proca action, and after that to the spontaneous symmetry breaking, which has completely different nature and properties from the explicit symmetry breaking.

\section{\label{secIV}Constrained $3BF$ action for the Proca field}

In order to study the electroweak theory, spontaneous symmetry breaking and the Higgs mechanism within the framework of higher gauge theory, an important step is to give a review of the Proca action written as a constrained $3BF$ theory, since the Higgs mechanism will naturally generate mass terms for the vector bosons. The $3BF$ formulation of the Proca action will therefore help us recognize these terms when we turn to the details of the Higgs mechanism.

In order to introduce the constrained $3BF$ action for the Proca field, the first step is to specify the choice of a $3$-group. The typical choice is the following. The three component Lie groups are given as:
\begin{equation} \label{eq:ProcaThreeGroup}
G = SO(3,1) \times SU(N) \,, \qquad H = \realni^4\,, \qquad L = \{ \one_L \}\,.
\end{equation}
This choice corresponds to the $SU(N)$ Yang-Mills field coupled to Einstein-Cartan gravity, with no scalar or fermion matter (since the group $L$ is trivial). The trivial choice of $L$ implies that the Peiffer lifting and the homomorphism $\delta$ are also trivial, as well as the action $\triangleright $ of the group $G$ onto $L$. What remains to be specified is the homomorphism $\partial$ and the action $\triangleright$ of the group $G$ onto itself and onto the group $H$. We choose the homomorphism $\partial$ to be trivial as well, while the action $\triangleright$ is specified as follows. The action of $G$ onto itself is given via the equations (\ref{eq:actionOfLorentzPartOfG}) and (\ref{eq:actionOfInternalPartOfG}), similar as for the Standard Model, while the action of $G$ onto $H$ is also given via the equations (\ref{eq:actionGOnH}).

In order to define the corresponding $3BF$ action, the symmetric nondegenerate invariant bilinear forms $\langle \_ \,, \_ \rangle_{\mathfrak{g}}$ and $\langle \_ \,, \_ \rangle_{\mathfrak{h}}$ are specified via the equations (\ref{snbfg}) and (\ref{snbfh}), respectively, while $\langle \_ \,, \_ \rangle_{\mathfrak{l}}$ is trivial. These choices simplify the $3BF$ action into a $2BF$ action, a special case of (\ref{eq:3BFforStandardModel}), given as:
\begin{equation}
S_{2BF}=\int B_{\alpha}\wedge F^{\alpha}+ B^{[ab]}\wedge R_{[ab]}+e_a\wedge\nabla\beta^{a}\,.
\end{equation}
Here the first term is the $BF$ term for the $SU(N)$ group corresponding to the Yang-Mills part, while the remaining two terms correspond to the gravitational part.

Once we have specified the topological part of the action, we deform it by adding appropriate constraints. In order to obtain appropriate dynamics for gravity, we have to add the gravitational constraint term (\ref{eq:gravConstraint}), while in order to obtain appropriate dynamics for the Yang-Mills field we similarly have to add the Yang-Mills constraint term (\ref{eq:YangMillsConstraintTerm}), and in this case one additional constraint, called the Proca constraint term:
\begin{equation} \label{eq:ProcaAction}
S = S_{2BF} + S_{\text{grav}} + S_{\text{Yang-Mills}} + S_{\text{Proca}} \,.
\end{equation}
The new Proca constraint term has the following form
\begin{equation} \label{eq:ProcaConstraintTerm}
S_{\text{Proca}} = \int \Theta^{\alpha ab}\wedge \left(\Xi_{\alpha abc}\varepsilon^{cdef}e_d\wedge e_e \wedge e_f+\frac{M}{g} \alpha_\alpha\wedge e_a \wedge e_b\right) + \frac{M}{g}\alpha^{\alpha}\wedge \Xi_{\alpha abc}e^a\wedge e^b\wedge e^c \,,
\end{equation}
where the $1$-form $\Theta^{\alpha ab}$ and $0$-form $\Xi_{\alpha abc}$ are new Lagrange multipliers, $M$ is the new parameter, while $g$ is the Yang-Mills coupling constant, corresponding to the choice of the coupling constant bilinear form in (\ref{eq:YangMillsConstraintTerm}) as:
\begin{equation} \label{eq:ProcaCouplingConstantMatrix}
C_{\alpha\beta} = \frac{1}{g^2} g_{\alpha\beta} \,.
\end{equation}

In order to demonstrate that the action (\ref{eq:ProcaAction}) really corresponds to the theory of the Proca field, we compute the corresponding equations of motion. Similarly to the case of the Standard Model, the variations of the action with respect to all fields will give the equations that can be solved for the multipliers,
\begin{equation} \label{eq:EoMsForAuxiliaryFields}
\begin{array}{rclcrclcrcl}
  M_{\alpha ab}&=&-\displaystyle\frac{1}{48}\varepsilon_{abcd} e^c{}_\mu e^d{}_\nu F_\alpha{}^{\mu\nu}\,,&\hspace*{0.5cm}&
  \lambda_{\alpha \mu\nu}&=&-F_{\alpha \mu\nu}\,,&\hspace*{0.5cm}&
  \zeta^{\alpha ab}&=&\displaystyle\frac{1}{4g^2}\varepsilon^{abcd} e_{c\mu} e_{d\nu} F^{\alpha\mu\nu}\,,\vphantom{\ds\int_a^b} \\

  \Theta^{\alpha ab}{}_\mu&=&\displaystyle\frac{M}{6g} \varepsilon^{abcd} \alpha^\alpha{}_\nu e_c{}^\nu e_{d\mu}\,,& &
  \lambda_{[ab]\mu\nu}&=&\ds R_{[ab]\mu\nu}\,,& &
  \Xi_{\alpha abc}&=&\displaystyle\frac{M}{6g} \varepsilon_{abcd} \alpha_{\alpha\mu} e^{d\mu}\,, \vphantom{\ds\int_a^b}  \\

  B_{\alpha \mu\nu}&=&-\displaystyle\frac{e}{2g^2} \varepsilon_{\mu\nu\rho\sigma} F_\alpha{}^{\rho\sigma} \,, & &
  \beta^a{}_{\mu\nu}&=&\ds 0\,, & & 
  B_{[ab]\mu\nu}&=&\ds \frac{1}{8\pi l_p^2}\varepsilon_{abcd}e^c{}_{\mu}e^d{}_{\nu}\,, \vphantom{\ds\int}  \\
\end{array}
\end{equation}
then the Einstein field equation (\ref{ajn}) for the stress-energy tensor of the form
\begin{equation} \label{eq:ProcaStressEnergy}
T_{\mu\nu}=\frac{1}{g^2}\left(F^\alpha{}_{\mu\rho}F_{\alpha\nu}{}^{\rho}-\frac{1}{4}g_{\mu\nu}F^\alpha{}_{\rho\sigma} F_\alpha{}^{\rho\sigma}\right)+\frac{M^2}{g^2}\left(\alpha_{\alpha\mu}\alpha^\alpha{}_{\nu}-\frac{1}{2}g_{\mu\nu}\alpha_\alpha{}^\rho\alpha^\alpha{}_{\rho}\right)\,,
\end{equation}
as well as the equations for the spin connection (\ref{spinskakoneksija}) and the torsion equation $T^a \equiv \nabla e^a = 0$, and finally the equation of motion for the vector boson field
\begin{equation} \label{eq:ProcaEoM}
    \nabla_{\mu} F^{\alpha \mu}{}_\nu -M^2\alpha^{\alpha}{}_\nu = 0\,,
\end{equation}
where $F^\alpha{}_{\mu\nu}$ is the standard Yang-Mills field strength tensor for the $SU(N)$ connection $\alpha^\alpha{}_\mu$\,. This is precisely the Proca equation for the field with mass $M$.

\red{In addition to the equations of motion, one can verify that the action (\ref{eq:ProcaAction}) corresponds to the Proca theory by eliminating all auxiliary fields. Since auxiliary fields are algebraically determined as functions of the dynamical fields, their equations of motion can be substituted back into the action, leading to the second-order formulation of the theory. In particular, substituting all equations (\ref{eq:EoMsForAuxiliaryFields}) into (\ref{eq:ProcaAction}), after a certain amount of straightforward algebra, one obtains precisely the traditional formulation of the action for the Proca field coupled to Einstein-Cartan gravity:
\begin{equation}
S = \int \frac{1}{16\pi l_p^2} \varepsilon^{abcd}\, R_{ab}\wedge e_c\wedge e_d - \frac{1}{g^2} F_{\alpha}\wedge \dual F^{\alpha} - \frac{1}{4!}\frac{M^2}{g^2} \alpha^\alpha{}_\mu \alpha_\alpha{}^\mu \, \lc^{abcd} \, e_a \wedge e_b \wedge e_c \wedge e_d\,.
\end{equation}
Here $\alpha_\alpha{}^\mu \equiv \alpha_{\alpha\nu} g^{\mu\nu}$ where $g^{\mu\nu} = \eta^{ab} e_a{}^\mu e_b{}^\nu$. Also, $\dual F$ denotes the Hodge dual of the $2$-form $F$:
\begin{equation} \label{eq:HodgeDualDef}
  \dual F^{\alpha}=\frac{1}{4}F^{\alpha}{}_{cd}\varepsilon^{abcd}e_a\wedge e_b\,.
\end{equation}
}

When dealing with the electroweak theory and the Standard Model, one encounters multiple Proca fields, with different masses $M$. In order to account for this, let us generalize the action (\ref{eq:ProcaAction}) to the case of multiple Proca fields. This is done by choosing a $3$-group with a modified group $G$ of the form:
\begin{equation} \label{eq:GeneralizedProcaGroup}
G = SO(3,1) \times \prod_i U(1) \times \prod_j SU(N_j)\,.
\end{equation}
Compared to (\ref{eq:ProcaThreeGroup}), one can see that the subgroup $SU(N)$ in $G$ has been substituted with multiple copies of $U(1)$ and $SU(N_j)$, depending on how many Proca fields we wish to have in the theory. The structure of the $3$-group remains essentially the same, in the sense that the action $\triangleright$ is extended from the $SU(N)$ case to the more general case in an obvious way, so that equations (\ref{eq:actionOfLorentzPartOfG}), (\ref{eq:actionOfInternalPartOfG}) and (\ref{eq:actionGOnH}) remain valid for the general choice (\ref{eq:GeneralizedProcaGroup}).

Given this more general choice of the $3$-group, the action for the theory formally still has the form (\ref{eq:ProcaAction}), but now the terms $S_{2BF}$ and $S_\text{Yang-Mills}$ correspond to the new choice of the internal gauge group, and the coupling constant bilinear form $C_{\alpha\beta}$ does not need to have the form (\ref{eq:ProcaCouplingConstantMatrix}) anymore, but instead it may depend on multiple coupling constants $g_i$, one for each term in the products in (\ref{eq:GeneralizedProcaGroup}). The only requirements on $C_{\alpha\beta}$ are that it must be symmetric, nondegenerate and $G$-invariant, since its eigenvalues should be $1/g_i^2$. Finally, the term $S_\text{Proca}$ becomes more complicated, and has the following form:
\begin{equation} \label{eq:GeneralProcaConstraintTerm}
    S_{\text{Proca}} = \int \Theta^{\alpha ab}\wedge \left(\Xi_{\alpha abc}\varepsilon^{cdef}e_d\wedge e_e \wedge e_f+ \vphantom{\frac{M}{g}} N_{\alpha\beta} \, \alpha^\beta \wedge e_a \wedge e_b\right) + \alpha^{\alpha}\wedge \tilde{N}_\alpha{}^\beta \Xi_{\beta abc}e^a\wedge e^b\wedge e^c \,,
\end{equation}
This constraint term features a new bilinear form $N_{\alpha\beta}$ and a new matrix $\tilde{N}_\alpha{}^\beta$, which are constant and arbitrary, representing new free parameters of the action. In order to understand their physical meaning, let us discuss the equations of motion for the action, as follows. First, the equations that can be solved for the multipliers are
\begin{equation} \label{eq:GeneralProcaMultiplierEoMs}
\begin{array}{rclcrclcrcl}
  M_{\alpha ab}&=&-\displaystyle\frac{1}{48}\varepsilon_{abcd} e^c{}_\mu e^d{}_\nu F_\alpha{}^{\mu\nu}\,,&\hspace*{0.5cm}&
  \lambda_{\alpha \mu\nu}&=&-F_{\alpha \mu\nu}\,,&\hspace*{0.5cm}&
  \zeta_\alpha{}^{ab}&=&\displaystyle\frac{1}{4}C_{\alpha\beta} \varepsilon^{abcd} e_{c\mu} e_{d\nu} F^{\beta\mu\nu}\,,\vphantom{\ds\int_a^b} \\

  \Theta^{\alpha ab}{}_\mu&=&\displaystyle\frac{1}{6} \tilde{N}_\beta{}^\alpha \varepsilon^{abcd} \alpha^\beta{}_\nu e_c{}^\nu e_{d\mu}\,,& &
  \lambda_{[ab]\mu\nu}&=&\ds R_{[ab]\mu\nu}\,,& &
  \Xi_{\alpha abc}&=&\displaystyle\frac{1}{6} N_{\alpha\beta} \varepsilon_{abcd} \alpha^\beta{}_\mu e^{d\mu}\,, \vphantom{\ds\int_a^b}  \\

  B_{\alpha \mu\nu}&=&-\displaystyle\frac{e}{2} C_{\alpha\beta} \varepsilon_{\mu\nu\rho\sigma} F^{\beta\rho\sigma} \,, & &
  \beta^a{}_{\mu\nu}&=&\ds 0\,, & & 
  B_{[ab]\mu\nu}&=&\ds \frac{1}{8\pi l_p^2}\varepsilon_{abcd}e^c{}_{\mu}e^d{}_{\nu}\,, \vphantom{\ds\int}  \\
\end{array}
\end{equation}
where we can see the presence of the parameters $C_{\alpha\beta}$, $N_{\alpha\beta}$ and $\tilde{N}_\alpha{}^\beta$. Next, the torsion equation $\nabla e^a = 0$ remains unchanged, while the equation of motion and the stress-energy tensor for the vector fields obtain the following form:
\begin{equation} \label{eq:GeneralProcaEoM}
    \nabla_{\mu} F^{\alpha \mu}{}_\nu -M^\alpha{}_\beta \alpha^{\beta}{}_\nu = 0\,,
\end{equation}
\begin{equation} \label{eq:GeneralProcaStressEnergy}
T_{\mu\nu}=C_{\alpha\beta}\left(F^\alpha{}_{\mu\rho}F^\beta{}_\nu{}^{\rho}-\frac{1}{4}g_{\mu\nu}F^\alpha{}_{\rho\sigma} F^{\beta\rho\sigma}\right)+C_{\alpha\beta} M^\beta{}_\gamma \left(\alpha^\alpha{}_\mu \alpha^\gamma{}_{\nu}-\frac{1}{2}g_{\mu\nu}\alpha^{\alpha\rho} \alpha^\gamma{}_{\rho}\right)\,.
\end{equation}
Here the new matrix $M^\alpha{}_\beta$ is constructed from $C_{\alpha\beta}$, $N_{\alpha\beta}$ and $\tilde{N}_\alpha{}^\beta$ as follows:
\begin{equation} \label{eq:SquaredMassMatrixDef}
M^\alpha{}_\beta = \frac{1}{2} \left( C^{-1} \right)^{\alpha\gamma} \left( \tilde{N}_\gamma{}^\delta N_{\delta\beta} + \tilde{N}_\beta{}^\delta N_{\delta\gamma} \right) \,.
\end{equation}
This matrix is interpreted as the squared-mass matrix of the theory. Note that due to the fact that $C_{\alpha\beta}$ is nondegenerate, it is also invertible. In order to interpret $M^\alpha{}_\beta$ as a matrix whose eigenvalues are squares of masses, the parameters $C_{\alpha\beta}$, $N_{\alpha\beta}$ and $\tilde{N}_\alpha{}^\beta$ have to be chosen so that (\ref{eq:SquaredMassMatrixDef}) is positive semi-definite. In such a case, choosing a basis in Lie algebra $\mathfrak{g}$ as an eigenbasis of $M^\alpha{}_\beta$, and denoting the respective eigenvalues as $M_{(\alpha)}^2$, one can rewrite the squared-mass matrix into the form
\begin{equation}
M^\alpha{}_\beta = M^2_{(\alpha)} \delta^\alpha_\beta \,,
\end{equation}
where the parentheses over the index $\alpha$ denote that this index is not summed over. Substituting this into the equation of motion (\ref{eq:GeneralProcaEoM}) we finally obtain
\begin{equation} \label{eq:GeneralProcaEoMfinal}
    \nabla_{\mu} F^{\alpha \mu}{}_\nu - M^2_{(\alpha)} \alpha^{\alpha}{}_\nu = 0\,.
\end{equation}
This is a set of equations of motion for several Proca fields, with (possibly different) masses $M_{(\alpha)}$, which explains why we can interpret $M^\alpha{}_\beta$ as the squared-mass matrix. Also, note that the obtained equation of motion (\ref{eq:GeneralProcaEoMfinal}) and the stress-energy tensor (\ref{eq:GeneralProcaStressEnergy}) are natural generalizations of their single Proca field counterparts (\ref{eq:ProcaEoM}) and (\ref{eq:ProcaStressEnergy}), respectively. \red{Moreover, similarly to the case of a single Proca field, one can substitute the algebraic equations of motion for the auxiliary fields (\ref{eq:GeneralProcaMultiplierEoMs}) back into the action (\ref{eq:ProcaAction}) with (\ref{eq:GeneralProcaConstraintTerm}) to obtain the traditional second-order formulation of the Proca theory coupled to Einstein-Cartan gravity.}

In order to be able to successfully compare, term by term, the Proca action with the action that will be obtained in Section \ref{secV} as a result of the Higgs mechanism, there is one more generalization that we need to do. In particular, we modify the Proca constraint (\ref{eq:GeneralProcaConstraintTerm}) by introducing two additional Lagrange multipliers, a $1$-form $\theta^\alpha$ and a $3$-form $\rho_\alpha$, as follows:
\begin{equation} \label{eq:ExtendedProcaConstraintTerm}
  S_{\text{Proca}} = \int \Theta^{\alpha ab}\wedge \left(\Xi_{\alpha abc}\varepsilon^{cdef}e_d\wedge e_e \wedge e_f+ \vphantom{\frac{M}{g}} N_{\alpha\beta}\, \alpha^\beta \wedge e_a \wedge e_b\right) + \alpha^{\alpha}\wedge \tilde{N}_\alpha{}^\beta \rho_\beta + \theta^\alpha \wedge \left( \rho_\alpha - \Xi_{\alpha abc}e^a\wedge e^b\wedge e^c \right) \,.
\end{equation}
The two additional Lagrange multipliers provide a convenient extension of the configuration space, so that it is compatible with the configuration space that will naturally appear in Section \ref{secV}. Other than that, the multipliers do not modify any other property of the Proca action. In particular, the equations of motion (\ref{eq:GeneralProcaMultiplierEoMs}) and (\ref{eq:GeneralProcaEoMfinal}), as well as the stress-energy tensor (\ref{eq:GeneralProcaStressEnergy}) and the torsion equation remain unchanged. Of course, extending the configuration space also means that we have two additional equations of motion, for the two new multipliers:
\begin{equation}
\theta^\alpha = - \tilde{N}_\beta{}^\alpha \alpha^\beta{}_\mu\,, \qquad
\rho_{\alpha \nu\rho\sigma} = e\, M_{\alpha\beta}\, \lc_{\mu\nu\rho\sigma} \alpha^{\beta\mu}\,.
\end{equation}
\red{As before, these two equations can also be readily substituted back into the action in order to obtain the traditional second-order formulation of the Proca action.}

This concludes our analysis of the higher gauge theory reformulation of the Proca action. The form of the terms in the Proca constraint (\ref{eq:ExtendedProcaConstraintTerm}) are precisely the type of terms one should look for in the Standard Model action after spontaneous symmetry breaking. As we shall see below, these kind of terms will be found precisely for the $W^\pm$ and $Z^0$ bosons in the electroweak theory.

\section{\label{secV}Spontaneous symmetry breaking and the Higgs mechanism}
\label{sec:sns}

The traditional formulation of the action for the Standard Model of elementary particles does not involve the $3BF$ action and simplicity constraint terms, but is rather expressed in the ordinary tensor form of the Lagrangian. One then performs a sequence of steps, comprising the Higgs mechanism, in order to rewrite the Lagrangian in the form where the full gauge symmetry is not manifest. The additional assumption that the vacuum state is not invariant with respect to the full gauge symmetry, but only one of its subgroups, and the corresponding gauge fixing of the Lagrangian, renders the gauge symmetry of the theory spontaneously broken.

In light of the framework of $3BF$ theory with constraints described in Section \ref{secII}, it is natural to ask whether the Higgs mechanism can be applied to the action (\ref{eq:RealisticAction}) which represents the Standard Model expressed in this new language. Answering that question is the topic of this Section.

\subsection{Constrained $3BF$ action for the electroweak theory}

In order to demonstrate the Higgs mechanism in the simplest way possible, let us restrict the action (\ref{eq:RealisticAction}) to the electroweak sector, and for the moment ignore the fermion spectrum. In other words, we choose the $3$-group in the following way:
\begin{equation} \label{eq:electroweakThreeGroup}
G=SO(3,1)\times SU(2)\times U(1)\,, \qquad H=\mathbb{R}^4\,, \qquad L=\mathbb{C}^4\,.
\end{equation}
The group $G$ features the Lorentz subgroup, the weak isospin $SU(2)$ subgroup, and the weak hypercharge $U(1)$ subgroup. The group $H$ remains the same as before, describing translations, while the group $L$ has been reduced to describe only the doublet of complex scalar fields. The homomorphisms $\delta$ and $\partial$ remain trivial, as well as the Peiffer lifting $\{\_\,,\_\}_{\text{pf}}$. Finally, the action of the group $G$ is defined as follows. It acts on itself via conjugation, the Lorentz part acts in the standard way onto the group $H$, and trivially onto the group $L$, thereby defining that all component fields from $L$ are scalar fields. The weak isospin and hypercharge act trivially on $H$, while they act in a nontrivial way on $L$. In order to explicitly state this action, it is useful to introduce the matrix notation for the generators $T_A$ of the group $L$, in an obvious way, as:
\begin{equation}
T_1 = \begin{pmatrix}
1\\
0\\
0\\
0
\end{pmatrix}
\,, \qquad
T_2 = \begin{pmatrix}
0\\
1\\
0\\
0
\end{pmatrix}
\,, \qquad
T_3 = \begin{pmatrix}
0\\
0\\
1\\
0
\end{pmatrix}
\,, \qquad
T_4 = \begin{pmatrix}
0\\
0\\
0\\
1
\end{pmatrix}
\,.
\end{equation}
Then, if we denote the generators of weak isospin as $\tau_i$ ($i=1,2,3$), and the generator of hypercharge as $\tau_0$, we have:
\begin{equation} \label{eq:actionOfTauOnT}
\tau_\alpha \triangleright T_A = \triangleright_{\alpha A}{}^B T_B\,,
\end{equation}
where the index $\alpha$ takes values $0,\dots, 3$, and thus conveniently counts all four generators $(\tau_0,\tau_i)$ of the group $SU(2)\times U(1)$. The coefficients are explicitly given as:
\begin{equation} \label{eq:TauOnTmatrices}
  \begin{array}{c}
    \triangleright_{0 A}{}^{B}=\ds\frac{i}{2}
    \begin{pmatrix}
    1 & 0 & 0 & 0\\
    0 & 1 & 0 & 0\\
    0 & 0 & -1 & 0\\
    0 & 0 & 0 & -1
    \end{pmatrix}
\,, \qquad
        \triangleright_{1 A}{}^{B}=\ds\frac{i}{2}
    \begin{pmatrix}
    0 & 1 & 0 & 0\\
    1 & 0 & 0 & 0\\
    0 & 0 & 0 & -1\\
    0 & 0 & -1 & 0
    \end{pmatrix}
    \,,
 \   \\
\\
    \triangleright_{2 A}{}^{B}=\ds\frac{i}{2}
    \begin{pmatrix}
    0 & i & 0 & 0\\
    -i & 0 & 0 & 0\\
    0 & 0 & 0 & i\\
    0 & 0 & -i & 0
    \end{pmatrix}
    \,, \qquad
    \triangleright_{3 A}{}^{B}=\ds\frac{i}{2}
    \begin{pmatrix}
    1 & 0 & 0 & 0\\
    0 & -1 & 0 & 0\\
    0 & 0 & -1 & 0\\
    0 & 0 & 0 & 1
    \end{pmatrix}
    \,.
    \end{array}
\end{equation}
Note also that the generators of $SU(2)\times U(1)$ satisfy the usual commutation relations
\begin{equation}
  f_{\alpha\beta\gamma}=\left\{
\begin{array}{ll}
  -\varepsilon_{\alpha\beta\gamma}\,, & \text{ for }\alpha,\beta,\gamma \neq 0\,, \\
  0 & \text{ otherwise.} \\
\end{array}
  \right.
\end{equation} 
This fixes the choice of the electroweak $3$-group. Next, the bilinear forms are defined in the natural way --- for the groups $G$ and $H$ they are defined as in (\ref{snbfg}) and (\ref{snbfh}), while for the group $L$ the choice may appear unusual:
\begin{equation} \label{eq:BilinearFormForHiggs}
        g_{AB}=\frac{1}{2}\begin{pmatrix}
        0 & 0 & 1 & 0\\
        0 & 0 & 0 & 1\\
        1 & 0 & 0 & 0\\
        0 & 1 & 0 & 0
        \end{pmatrix}\,.
\end{equation}
This is only apparent, since we wish to represent an element of the algebra $\mathfrak{l}$ in the form
\begin{equation} \label{eq:PhiComponentsInNaturalBasis}
\phi \equiv \phi^A T_A \equiv \phi_+ T_1 + \phi_0 T_2 + \phi_+^\dagger T_3 + \phi_0^\dagger T_4 = \begin{pmatrix}
\phi_+\\
\phi_0\\
\phi_+^{\dagger}\\
\phi_0^{\dagger}
\end{pmatrix}\,.
\end{equation}
However, if one switches to a new basis in $\mathfrak{l}$ as
\begin{equation} \label{eq:ChangeOfBasisInL}
\tilde{T}_1 = T_1 + T_3 \,, \qquad
\tilde{T}_2 = iT_1 -i T_3 \,, \qquad
\tilde{T}_3 = T_2 + T_4 \,, \qquad
\tilde{T}_4 = iT_2 -iT_4 \,,
\end{equation}
the same algebra element can be rewritten as
\begin{equation} \label{eq:elementInTildeTbasis}
\phi = \phi_1 \tilde{T_1} + \phi_2 \tilde{T_2} + \phi_3 \tilde{T_3} + \phi_4 \tilde{T_4} \,, 
\end{equation}
where $\phi_1,\dots,\phi_4$ are real-valued components, and there is a natural correspondence between the coefficients:
\begin{equation}
\phi_+ = \phi_1 + i \phi_2\,, \qquad \phi_0 = \phi_3 + i \phi_4 \,, \qquad
\phi_+^\dagger = \phi_1 - i \phi_2\,, \qquad \phi_0^\dagger = \phi_3 - i \phi_4 \,.
\end{equation}
In the basis $\tilde{T}_A$ the bilinear form (\ref{eq:BilinearFormForHiggs}) becomes the unit diagonal matrix. The basis $\tilde{T}_A$ is convenient because of the diagonal bilinear form and the real-valued components, while the basis $T_A$ is convenient because it is an eigenbasis for the weak isospin and weak hypercharge opeators (and as we shall see, also for the electromagnetic charge operator). We will be frequently switching between these two bases throughout this Section.

We should also note that (\ref{eq:elementInTildeTbasis}) can be understood as an element of the $4$-dimensional real-valued Lie algebra $L = \realni^4$, or equivalently of the $2$-dimensional complex-valued Lie algebra $L = \kompleksni^2$ (which is implicitly being used in most standard textbooks dealing with the Higgs mechanism). On the other hand, (\ref{eq:PhiComponentsInNaturalBasis}) is an element of the $4$-dimensional complex-valued Lie algebra $L = \kompleksni^4$, which is a complexification of $\realni^4$, and if we wish to be able to seamlessly switch from the basis $T_A$ to $\tilde{T}_A$ and back, it is far more convenient to work with the complexified algebra. Hence the choice $L = \kompleksni^4$ in the electroweak $3$-group (\ref{eq:electroweakThreeGroup}).

Once we have specified the choice of the $3$-group and the choices for the bilinear forms, the action for the electroweak theory can be written as:
\begin{equation} \label{eq:dej} 
S=S_{3BF}+S_{\text{grav}}+S_{\text{scal}}+S_{\text{Yang-Mills}}+S_{\text{Higgs}}+S_{\text{CC}}\,.
\end{equation}
It is similar in form to (\ref{eq:RealisticAction}), where the constraint terms related to fermions have been ommited. The coupling constant bilinear form in $S_{\text{Yang-Mills}}$ is given as
\begin{equation} \label{eq:CouplingConstantMatrix}
{C}_{\alpha\beta}=\begin{pmatrix}
    \frac{1}{g_0^2} & 0 & 0 & 0\\
    0 & \frac{1}{g_1^2} & 0 & 0\\
    0 & 0 & \frac{1}{g_1^2} & 0\\
    0 & 0 & 0 & \frac{1}{g_1^2}
    \end{pmatrix}\,,
\end{equation}
reflecting the structure of the $SU(2)\times U(1)$ group.

\subsection{Overview of the Higgs mechanism}

There are three main steps in the Higgs mechanism:
\begin{itemize}
\item discussion of the stable vacuum,
\item introduction of a change of variables,
\item gauge fixing of the scalar fields.
\end{itemize}
In order to understand the details of the Higgs mechanism in the framework of the action (\ref{eq:dej}), it is illustrative to repeat these main steps using the new variables and notation.

The analysis of the stable vacuum is essentially identical to the usual case of the Higgs mechanism. The $S_\text{Higgs}$ constraint introduces the following potential for the scalar field,
\begin{equation} \label{eq:HiggsPotential}
V(\phi)=2\chi\left(\phi^A\phi_A-v^2\right)^2\,,
\end{equation}
and one can observe that the stable vacuum is not unique, but is represented by a 3-sphere of points $\phi^A\phi_A = v^2$ in the configuation space. In order to rewrite the action in terms of fields that become equal to zero at some given vacuum point, one is led to introduce a change of variables from $(\phi_1, \phi_2, \phi_3, \phi_4)$ to $(\phi_1, \phi_2, h, \phi_4)$, where $h(x)$ is the new scalar field, obrained by translating $\phi_3$ by $v$:
\begin{equation}
\phi_3(x) = v + h(x)\,.
\end{equation}
This corresponds to the point $(0,0,v,0)$ on the 3-sphere as our vacuum of choice, by convention. Of course, this convention is completely arbitrary, and nothing in the rest of the analysis depends on this choice. The change of variables is given in terms of the basis $\tilde{T}_A$, while in terms of our original basis $T_A$ we have:
\begin{equation}
\phi^A=\begin{pmatrix}
\phi_+\\
\phi_0\\
\phi_+^{\dagger}\\
\phi_0^{\dagger}
\end{pmatrix}=\begin{pmatrix}
\phi_1+i\phi_2\\
v+h+i\phi_4\\
\phi_1-i\phi_2\\
v+h-i\phi_4
\end{pmatrix}\,.
\end{equation}
Finally, given this relation, one can observe that the components $\phi_1$, $\phi_2$ and $\phi_4$ are in fact equivalent (up to linear order) to three gauge parameters of the $\mathfrak{g}$-gauge transformation
\begin{equation} \label{eq:PhiGaugeTransformation}
\phi \to \phi' = e^{\xi^\alpha \tau_\alpha} \triangleright \phi\,.
\end{equation}
Namely, using the action (\ref{eq:actionOfTauOnT}) of the generators of algebra $\mathfrak{g}$ on the generators of algebra $\mathfrak{l}$, one can start from the following state and the choice of the following gauge parameters,
\begin{equation} \label{eq:GaugeParametersAndPhis}
\phi^A = \begin{pmatrix}
0\\
v+h\\
0\\
v+h\\
\end{pmatrix}\,,
\qquad
\xi^{\alpha}(\phi^A)=\frac{1}{v}\begin{pmatrix}
    \phi_4\\
    2\phi_2\\
    2\phi_1\\
    -\phi_4
    \end{pmatrix} + \cO(\phi^2)\,,
\end{equation}
and evaluate gauge transformation (\ref{eq:PhiGaugeTransformation}) on the above state to obtain:
\begin{equation}
\phi'{}^A = e^{\xi^\alpha(\phi)\tau_\alpha}\begin{pmatrix}
0\\
v+h\\
0\\
v+h\\
\end{pmatrix}
=\begin{pmatrix}
\phi_1+i\phi_2\\
v+h+i\phi_4\\
\phi_1-i\phi_2\\
v+h-i\phi_4
\end{pmatrix}\,.
\end{equation}
Therefore, we see that the fields $\phi_1$, $\phi_2$ and $\phi_4$ can be understood as gauge degrees of freedom, given by the relation (\ref{eq:GaugeParametersAndPhis}). One can conclude that only the field $h$ is physical, since it cannot be removed by a $\mathfrak{g}$-gauge transformation.

Next, one can see that, even after the removal of $\phi_1$, $\phi_2$ and $\phi_4$ by using a gauge transformation, the state $\phi^A$ in (\ref{eq:GaugeParametersAndPhis}) still remains invariant with respect to a $U(1)$ subgroup of $G$. Denoting the generator of this stabilizer group as $Q$, one can easily see from (\ref{eq:TauOnTmatrices}) that the stabilizer requirement $Q\triangleright\phi = 0$ is satisfied for
\begin{equation} \label{eq:StabilizerDef}
Q = \tau_0 + \tau_3\,.
\end{equation}
This equation is known by the name Gell-Mann--Nishijima formula (for electroweak interactions). Phenomenologically, $Q$ corresponds to the electromagnetic charge $q$, specifically $q$ is an eigenvalue of the operator $-iQ$, and the corresponding $U(1)$ stabilizer group is the gauge group of electrodynamics. From the stabilizer requirement one can observe that the Higgs field $h(x)$ has no electric charge, since it corresponds for the eigenvalue $q=0$.

Let us note that the above results do not depend in any way on the choice of the vacuum point $(0,0,v,0)$ on the 3-sphere. One could have chosen any other point, in which case the only difference is that the solution of the stabilizer equation $Q\triangleright \phi = 0$ would be slighly more general:
\begin{equation}
Q = \tau_0 + \vec\alpha \cdot \vec\tau\,, \qquad \vec\alpha \in \realni^3, \qquad \| \vec\alpha \|^2 = 1\,.
\end{equation}
Here $\vec\tau$ is understood as a triple $(\tau_1,\tau_2,\tau_3)$. In particular, the electromagnetic charge of the Higgs field would remain zero even in this case.

\subsection{Transformation of the action}

Let us now turn to the problem of the transformation of the action with respect to the gauge transformation of the scalar field that removes the components $\phi_1$, $\phi_2$ and $\phi_4$,
\begin{equation} \label{eq:TransformationRuleForPhi}
\phi^A \to \left(e^{-\xi} \triangleright \phi\right)^A = \begin{pmatrix}
0\\
v+h\\
0\\
v+h\\
\end{pmatrix}\,,
\end{equation}
where $\xi \equiv \xi^\alpha\tau_\alpha $, and the parameters $\xi^\alpha$ are given in (\ref{eq:GaugeParametersAndPhis}). In order to see what happens to the action (\ref{eq:dej}), let us first note that the remaining variables that enter the action transform as follows. The transformation of the 3-connection variables $(\alpha,\omega,\beta,\tilde{\gamma})$ is given as:
\begin{equation}
\alpha'=e^{-\xi}\left(\alpha+\rmd \right)e^{\xi}\,, \qquad
\omega' = \omega \,, \qquad
\beta'=\beta\,, \qquad
\tilde{\gamma}'=e^{-\xi}\triangleright\tilde{\gamma}\,.
\end{equation}
The corresponding curvatures transform as:
\begin{equation}
F'=e^{-\xi}Fe^{\xi}\,, \qquad
R'=R\,, \qquad
\cG'=\cG\,, \qquad
\cH'=e^{-\xi} \triangleright \cH\,.
\end{equation}
The transformations of the Lagrange multipliers which appear in the topological sector of the action, namely $B_\alpha$, $B_{[ab]}$, $e_a$ and $\phi^A$, are given as:
\begin{equation}
  B'_\alpha = \left( e^{-\xi} B e^\xi \right)_\alpha \,, \qquad
  B'_{[ab]}=B_{[ab]} \,, \qquad
  e'_a=e_a\,, \qquad
\end{equation}
while the transformation of $\phi^A$ is already spelled out in (\ref{eq:TransformationRuleForPhi}). Next, the $\mathfrak{g}$-valued Lagrange multipliers which appear in the constraint sector of the action, namely $\lambda_\alpha$, $\lambda_{[ab]}$, $M_{\alpha ab}$ and $\zeta_{\alpha ab}$, transform as:
\begin{equation}
\lambda'_\alpha=\left(e^{-\xi}\lambda e^{\xi}\right)_\alpha\,, \qquad
\lambda'_{[ab]}=\lambda_{[ab]}\,, \qquad
M'_{\alpha ab}=\left( e^{-\xi} M e^{\xi}\right)_{\alpha ab}\,, \qquad
\zeta'_{\alpha ab}=\left( e^{-\xi} \zeta e^{\xi}\right)_{\alpha ab} \,.
\end{equation}
The $\mathfrak{l}$-valued Lagrange multipliers which appear in the constraint sector of the action, namely $\tilde{\lambda}_A$, $\Lambda_{abA}$ and $H_{abcA}$, transform as:
\begin{equation}
\tilde{\lambda}'_A=\left( e^{-\xi}\triangleright\tilde{\lambda} \right)_A\,, \qquad
\Lambda'_{abA}=\left(e^{-\xi}\triangleright\Lambda\right)_{abA}\,, \qquad
H'_{abcA}=\left(e^{-\xi}\triangleright H\right)_{abcA}\,.
\end{equation}
Finally, the constraint part of the action also features the covariant derivative $\nabla\phi$, which transforms in a covariant way,
\begin{equation}
  \left(\nabla\phi\right)'= e^{-\xi}\triangleright\left(\nabla\phi\right)\,,
\end{equation}
as expected for a covariant derivative.

In addition to all of the above fields, the action also features the bilinear form of coupling constants, $C_{\alpha\beta}$, given by (\ref{eq:CouplingConstantMatrix}). One can observe that this bilinear form is in fact term-by-term proportional to the already introduced bilinear form $\langle\_\,,\_\rangle_{\mathfrak{g}}$, as follows:
\begin{equation}
C_{\alpha\beta} = \cC(\tau_\alpha, \tau_\beta) \equiv \frac{\delta^j_\alpha \delta^k_\beta}{g_1^2} \langle \tau_j, \tau_k \rangle_{\mathfrak{g}} + \frac{\delta^0_\alpha \delta^0_\beta}{g_0^2} \langle \tau_0, \tau_0 \rangle_{\mathfrak{g}}\,.
\end{equation}
The two terms in the sum correspond to bilinear forms $\langle\_\,,\_\rangle_{\mathfrak{su}(2)}$ and $\langle\_\,,\_\rangle_{\mathfrak{u}(1)}$, respectively. Given that the gauge transformation can be represented in the form $e^{-\xi^i \tau_i} \times e^{-\xi^0\tau_0}$, owing to the direct product structure in the group $SU(2)\times U(1)$, each term in the gauge transformation leaves the corresponding bilinear form invariant,
\begin{equation}
\langle e^{-\xi^i \tau_i}\triangleright \tau_j , e^{-\xi^i \tau_i}\triangleright \tau_k \rangle_{\mathfrak{su}(2)} = \langle \tau_j,\tau_k\rangle_{\mathfrak{su}(2)}\,, \qquad
\langle e^{-\xi^0 \tau_0}\triangleright \tau_0 , e^{-\xi^0 \tau_0}\triangleright \tau_0 \rangle_{\mathfrak{u}(1)} = \langle \tau_0,\tau_0\rangle_{\mathfrak{u}(1)}\,,
\end{equation}
as a consequence of the postulated $G$-invariance property of the bilinear form $\langle\_\,,\_\rangle_{\mathfrak{g}}$. This renders the bilinear form of coupling constants gauge invariant:
\begin{equation}
C'{}_{\alpha\beta} = C_{\alpha\beta} \,.
\end{equation}

At this point we are ready to discuss the transformation of the action with respect to (\ref{eq:TransformationRuleForPhi}). Namely, the action (\ref{eq:dej}) is a functional of all fields mentioned above,
\begin{equation}
\alpha^\alpha,\;\omega^{[ab]},\;\beta^a,\;\tilde{\gamma}^A, \;B_\alpha,\; B_{[ab]},\; e_a ,\; \lambda_\alpha,\; \lambda_{[ab]},\; M_{\alpha ab} ,\; \zeta_{\alpha ab} ,\; \tilde{\lambda}_A,\; \Lambda_{abA},\; H_{abcA}, \; \phi^A ,
\end{equation}
or in other words, the fields in the above list define a kinematical configuration space of our action. However, not every term in the action is a function of $\phi^A$ in particular. Therefore, when performing the gauge transformation (\ref{eq:TransformationRuleForPhi}), terms independent of $\phi^A$ will remain the same, while the terms dependent of $\phi^A$ will transform in a nontrivial way, reducing the full configuration space to a smaller one, defined by the fields
\begin{equation} \label{eq:ReducedConfSpace}
\alpha^\alpha,\;\omega^{[ab]},\;\beta^a,\;\tilde{\gamma}^A, \;B_\alpha,\; B_{[ab]},\; e_a ,\; \lambda_\alpha,\; \lambda_{[ab]},\; M_{\alpha ab} ,\; \zeta_{\alpha ab} ,\; \tilde{\lambda}_A,\; \Lambda_{abA},\; H_{abcA} ,\; h ,
\end{equation}
which differ from the original set in the replacement $(\phi^1,\phi^2,\phi^3,\phi^4) \to (0,0,v+h,0)$. The task is then to determine the form of the action $\tilde S$ which is defined on this reduced configuration space, schematically defined by the transformation:
\begin{equation} \label{eq:SchematicTransfOfAction}
S[\dots, \phi^A] \quad \xrightarrow{\quad e^{-\xi} \quad} \quad \tilde{S}[\dots,h] \equiv S[\dots,\phi^A] \left.\vphantom{\ds\int} \right\vert_{\substack{\phi^1=\phi^2=\phi^4=0 \\ \phi^3=v+h}}\,.
\end{equation}
One can immediately observe that $S_\text{grav}$, $S_\text{Yang-Mills}$ and $S_\text{CC}$ transform in a trivial way, since they do not depend on $\phi$:
\begin{equation} \label{eq:TransfOfTrivialActions}
S_\text{grav} \;\; \xrightarrow{e^{-\xi}} \;\; \tilde{S}_\text{grav} = S_\text{grav} \,, \qquad
S_\text{Yang-Mills} \;\; \xrightarrow{e^{-\xi}} \;\; \tilde{S}_\text{Yang-Mills} = S_\text{Yang-Mills} \,, \qquad
S_\text{CC} \;\; \xrightarrow{e^{-\xi}} \;\; \tilde{S}_\text{CC} = S_\text{CC} \,.
\end{equation}
Moreover, the $2BF$ part of $S_{3BF}$ also transforms trivially, for the same reason.

On the other hand, the third term in $S_{3BF}$, as well as $S_\text{scal}$ and $S_\text{Higgs}$ require more attention. Let us discuss first the $S_\text{Higgs}$ term. Specifically, we have that
\begin{equation}
S_\text{Higgs} = - \int \frac{1}{4!} V(\phi) \, \lc_{abcd} e^a \wedge e^b \wedge e^c \wedge e^d\,,
\end{equation}
where under the transformation (\ref{eq:SchematicTransfOfAction}) the Higgs potential $V(\phi)$ (see (\ref{eq:HiggsPotential})) becomes
\begin{equation} \label{eq:TransfOfHiggsPotential}
V(\phi) \;\; \xrightarrow{e^{-\xi}} \;\; V(h)\equiv 8v^2\chi h^2+8v\chi h^3 + 2\chi h^4 \,.
\end{equation}
Therefore, we see that
\begin{equation} \label{eq:TransfOfHiggsAction}
S_\text{Higgs} \;\; \xrightarrow{e^{-\xi}} \;\; \tilde{S}_\text{Higgs} = - \int \frac{1}{4!} V(h) \, \lc_{abcd} e^a \wedge e^b \wedge e^c \wedge e^d\,.
\end{equation}
From the form of the quadratic term in the resulting potential (\ref{eq:TransfOfHiggsPotential}) and the general form of the mass term for a single real scalar field (\ref{skalarna_masa}), one can read off the value of the Higgs mass as:
\begin{equation}
m = 2 v \sqrt{2 \chi} \,.
\end{equation}
Let us also note here that one could choose a potential which has a form alternative to (\ref{eq:HiggsPotential}), for example
\begin{equation}
V_\text{alt}(\phi) = 2\chi \left( \phi^A \phi_A \right)^2 - 4\chi v^2 \phi^A \phi_A\,.
\end{equation}
This potential differs from (\ref{eq:HiggsPotential}) by a constant term $2\chi v^4$, which would then combine with $S_\text{CC}$ to give a different value of the cosmological constant. However, the potential (\ref{eq:HiggsPotential}) does not suffer from this problem, and in our case the CC term of the action remains the same before and after spontaneous symmetry breaking.

Next let us discuss the $S_{3BF}$ term. Note that only the final term in  $S_{3BF}$ depends on $\phi$, while the remainder does not and can be denoted as $S_{2BF}$. Then, using a suitable change of basis $T_A \to \tilde{T}_A$ in the Lie algebra $\mathfrak{l}$ (see (\ref{eq:ChangeOfBasisInL})), with an additional notation for the indices $A\to (\bar{A},H)$ where $\bar{A} \in \{ 1,2,4 \}$ and $H\equiv 3$, we have
\begin{equation} \label{eq:ActionCalculation1}
\begin{array}{ccl}
  S_{3BF} = \ds S_{2BF} + \int \phi^A\nabla\tilde{\gamma}_A & \xrightarrow{e^{-\xi}} & \ds S_{2BF} + \int (v+h)\left( (\nabla \tilde{\gamma})_0 + (\nabla\tilde{\gamma})_{0^\dag} \right) \vphantom{\ds\int} \\
  & & \ds = S_{2BF} + \int (v+h) \left( \rmd \tilde{\gamma}_H + \triangleright{}_\alpha{}^{\bar{A}}{}_H \, \alpha^\alpha \wedge \tilde{\gamma}_{\bar{A}}  \right) \vphantom{\ds\int} \\
  & & \ds = S_{2BF} + \int h \rmd \tilde{\gamma}_H + v \rmd \tilde{\gamma}_H + (v+h) \triangleright{}_\alpha{}^{\bar{A}}{}_H \, \alpha^\alpha \wedge \tilde{\gamma}_{\bar{A}}  \vphantom{\ds\int} \\
  & & \ds = \tilde{S}_{3BF} + \int v \rmd \tilde{\gamma}_H + (v+h) \triangleright{}_\alpha{}^{\bar{A}}{}_H \, \alpha^\alpha \wedge \tilde{\gamma}_{\bar{A}} \,, \vphantom{\ds\int} \\
\end{array}
\end{equation}
where the new action $\tilde{S}_{3BF}$ is defined as a functional over the reduced configuration space (\ref{eq:ReducedConfSpace}) as
\begin{equation} \label{eq:ActionCalculation2}
\tilde{S}_{3BF} = S_{2BF} + \int h \rmd \tilde{\gamma}_H\,.
\end{equation}
In Section \ref{secVI} we shall discuss in detail its corresponding $3$-group. Therefore, we conclude that
\begin{equation} \label{eq:TransfOf3BFAction}
S_{3BF} \;\; \xrightarrow{e^{-\xi}} \;\; \tilde{S}_{3BF} + \int v \rmd \tilde{\gamma}_H + (v+h) \triangleright{}_\alpha{}^{\bar{A}}{}_H \, \alpha^\alpha \wedge \tilde{\gamma}_{\bar{A}}\,,
\end{equation}
where the extra terms will later be grouped together with extra terms from other parts of the action and discussed in detail.

In equations (\ref{eq:ActionCalculation1}), (\ref{eq:ActionCalculation2}) and (\ref{eq:TransfOf3BFAction}) we have made use of the basis (\ref{eq:ChangeOfBasisInL}) in the Lie algebra $\mathfrak{l}$, so that we can introduce $\tilde{\gamma}_H \equiv \tilde{\gamma}_0 + \tilde{\gamma}_{0^\dag}$. The action $\triangleright$ was represented via the matrices (\ref{eq:TauOnTmatrices}) in the original basis $T_A$, while in this basis it is now broken into the following set of components:
\begin{equation}
\triangleright{}_{\alpha H}{}^{\bar{A}}\,, \qquad 
\triangleright{}_{\alpha \bar{A}}{}^H\,, \qquad 
\triangleright{}_{\alpha \bar{A}}{}^{\bar{B}}\,, \qquad
\triangleright{}_{\alpha H}{}^H\,.
\end{equation}
Since in this basis the bilinear form $g_{AB}$ is diagonal, in fact $g_{AB} = \delta_{AB}$, a consequence of the Theorem from Section \ref{secII} is that all these components have vanishing diagonal elements, in particular $\triangleright{}_{\alpha H}{}^H = 0$, which implies that $\nabla \tilde{\gamma}_H \equiv \rmd \tilde{\gamma}_H$ and justifies the identification (\ref{eq:ActionCalculation2}). Moreover, the components $\triangleright{}_{\alpha \bar{A}}{}^{\bar{B}}$ drop out of equations (\ref{eq:ActionCalculation1}), (\ref{eq:ActionCalculation2}) and (\ref{eq:TransfOf3BFAction}) and do not appear anywhere. This leaves us with the remaining set of relevant components, which can be represented in matrix form as follows:
\begin{equation} \label{eq:TriangleMatricesInNewBasis}
\triangleright{}_{\alpha H}{}^{\bar{A}} = \frac{1}{2}
\begin{pmatrix}
  0 & 0 & 1 \\
  0 & 1 & 0 \\
  1 & 0 & 0 \\
  0 & 0 & -1
\end{pmatrix}
\,, \qquad
\triangleright{}_{\alpha \bar{A}}{}^H =  \frac{1}{2}
\begin{pmatrix}
  0 & 0 & -1 \\
  0 & -1 & 0 \\
  -1 & 0 & 0 \\
  0 & 0 & 1
\end{pmatrix}
\,.
\end{equation}
Note that here $\alpha$ is the row index and $\bar{A}$ is the column index, while $H\equiv 3$ is constant.

Finally, let us discuss the $S_\text{scal}$ term, which originally has the form
\begin{equation}
  S_\text{scal} = \int \tilde{\lambda}^A\wedge\left(\tilde{\gamma}_{A}-H_{abcA}e^a\wedge e^b\wedge e^c\right) + \Lambda^{abA}\wedge H_{abcA}\varepsilon^{cdef}e_d\wedge e_e\wedge e_f -
\Lambda^{abA} \wedge \left(\nabla \phi \right)_A \wedge e_a\wedge e_b \,,
\end{equation}
and similarly to $S_{3BF}$, it also depends on $\phi^A$ only in the final term, while the remainder is independent of $\phi^A$. Splitting the index $A$ into $(\bar{A},H)$, the constraint transforms into:
\begin{equation}
\begin{array}{ccl}
  S_\text{scal} & \xrightarrow{e^{-\xi}} & \ds \int \tilde{\lambda}^H\wedge\left(\tilde{\gamma}_H-H_{abcH}e^a\wedge e^b\wedge e^c\right) + \Lambda^{abH}\wedge H_{abcH}\varepsilon^{cdef}e_d\wedge e_e\wedge e_f -
\Lambda^{abH}\wedge \rmd h \wedge e_a\wedge e_b
  \vphantom{\ds\int} \\
   & & + \ds \tilde{\lambda}^{\bar{A}}\wedge\left(\tilde{\gamma}_{\bar{A}}-H_{abc\bar{A}}e^a\wedge e^b\wedge e^c\right) + \Lambda^{ab\bar{A}}\wedge H_{abc\bar{A}}\varepsilon^{cdef}e_d\wedge e_e\wedge e_f -
  \Lambda^{ab\bar{A}}\wedge \alpha^{\alpha} \triangleright_{\alpha}{}^H{}_{\bar{A}} (v+h)\wedge e_a\wedge e_b\,.
  \vphantom{\ds\int} \\
\end{array}
\end{equation}
Note that the terms in the first row on the right-hand side are precisely the terms that define the scalar constraint for a single real scalar field, as a functional over the reduced configuration space (\ref{eq:ReducedConfSpace}). Denoting those terms as $\tilde{S}_\text{scal}$, we conclude that
\begin{equation} \label{eq:TransfOfScalAction}
\begin{array}{ccl}
  S_\text{scal} & \xrightarrow{e^{-\xi}} & \ds \tilde{S}_\text{scal} + \int \tilde{\lambda}^{\bar{A}}\wedge\left(\tilde{\gamma}_{\bar{A}}-H_{abc\bar{A}}e^a\wedge e^b\wedge e^c\right)
  \vphantom{\ds\int} \\
   & & \ds \hphantom{mmmm} + \Lambda^{ab\bar{A}}\wedge H_{abc\bar{A}}\varepsilon^{cdef}e_d\wedge e_e\wedge e_f
  \vphantom{\ds\int} \\
   & & \ds \hphantom{mmmm} - \Lambda^{ab\bar{A}}\wedge \alpha^{\alpha} \triangleright_{\alpha}{}^H{}_{\bar{A}} (v+h)\wedge e_a\wedge e_b\,,
  \vphantom{\ds\int} \\
\end{array}
\end{equation}
where we again have three extra terms which will be grouped together with the remainder of the action.

After we have discussed all parts of the action (\ref{eq:dej}) in the context of the transformation (\ref{eq:SchematicTransfOfAction}), we can put all the pieces together, and compare the full action with the actions for the Proca and massive scalar fields. However, in order to make this comparison more transparent, it is useful to introduce yet some more notation. In particular, let us introduce a bilinear form $\kappa^{\alpha\beta}$ so that it satisfies the following identity:
\begin{equation} \label{eq:KappaIdentity}
\kappa^{\alpha\beta}\triangleright_{\alpha H}{}^{\bar{A}}\triangleright_{\beta \bar{B}}{}^H=-\frac{1}{4}\delta^{\bar{A}}_{\bar{B}}\,.
\end{equation}
This bilinear form is not unique. Namely, since the matrices (\ref{eq:TriangleMatricesInNewBasis}) are of rank $3$, there exists a projector $P_\alpha{}^\beta$ which satisfies
\begin{equation} \label{eq:ProjectorDef}
P_\alpha{}^\beta P_\beta{}^\gamma = P_\alpha{}^\gamma\,, \qquad
P_\alpha{}^\alpha = 3\,, \qquad
P_{\alpha\beta} = P_{\beta\alpha}\,, \qquad
P_\alpha{}^\beta \triangleright_{\beta H}{}^{\bar{A}} = \triangleright_{\alpha H}{}^{\bar{A}}\,.
\end{equation}
Note that a projector that satisfies (\ref{eq:ProjectorDef}) also satisfies the identity $P_\alpha{}^\beta \triangleright_{\beta \bar{A}}{}^H = \triangleright_{\alpha \bar{A}}{}^H $, since $ \triangleright_{\alpha \bar{A}}{}^H$ has the same components as $ \triangleright_{\alpha H}{}^{\bar{A}} $ up to an overall minus sign, see (\ref{eq:TriangleMatricesInNewBasis}).
Therefore, the bilinear form $\kappa^{\alpha\beta}$ is defined up to a term of the form
\begin{equation} \label{eq:ArbitrarinessOfKappaMatrix}
\kappa^{\alpha\beta} \to \kappa^{\alpha\beta} + \left[ \delta^{(\alpha}_\gamma - P_\gamma{}^{(\alpha}  \right] A^{\beta)\gamma}\,,
\end{equation}
where $A^{\alpha\beta}$ is an arbitrary matrix, while the parentheses on the indices denote symmetrization. One can recognize that the term in the brackets is the orthogonal projector, which maps into the kernel of the matrices (\ref{eq:TriangleMatricesInNewBasis}). This arbitrariness guarantees that the bilinear form $\kappa^{\alpha\beta}$ can be chosen to be invertible.
The projector $P_\alpha{}^\beta$ can be explicitly evaluated using the definition (\ref{eq:ProjectorDef}) and the matrices (\ref{eq:TriangleMatricesInNewBasis}), while one convenient choice of the bilinear form $\kappa^{\alpha\beta}$ can be evaluated from (\ref{eq:KappaIdentity}), so that they can be written in matrix form as follows:
\begin{equation} \label{eq:ProjectorAndKappaMatrices}
  P_\alpha{}^\beta = \frac{1}{2} \begin{pmatrix}
    1 & 0 & 0 & -1 \\
    0 & 2 & 0 & 0 \\
    0 & 0 & 2 & 0 \\
    -1 & 0 & 0 & 1
\end{pmatrix}\,, \qquad
  \kappa^{\alpha\beta} = \frac{1}{2} \begin{pmatrix}
    1 & 0 & 0 & 0 \\
    0 & 2 & 0 & 0 \\
    0 & 0 & 2 & 0 \\
    0 & 0 & 0 & 1
\end{pmatrix}\,.
\end{equation}
In addition to the projector and $\kappa^{\alpha\beta}$, we can now also introduce the following quantities:
\begin{equation} \label{eq:Hdefinitions}
\theta^{\alpha} \equiv -2 \kappa^{\alpha\beta} \triangleright_{\beta}{}^H{}_{\bar{A}}\tilde{\lambda}^{\bar{A}}\,,\qquad
\Theta^{\alpha ab} \equiv -2 \kappa^{\alpha\beta} \triangleright_{\beta}{}^H{}_{\bar{A}}\Lambda^{ab\bar{A}}\,,\qquad
\rho_{\alpha} \equiv 2\triangleright_{\alpha}{}^{\bar{A}}{}_{H} \tilde{\gamma}_{\bar{A}}\,, \qquad
\Xi_{\alpha abc} \equiv 2\triangleright_{\alpha}{}^{\bar{A}}{}_{H}H_{abc\bar{A}}\,.
\end{equation}
These new quantities satisfy four fundamental identities,
\begin{equation} \label{eq:Hidentities}
\theta^{\alpha } \wedge \rho_{\alpha}=\tilde{\lambda}^{\bar{A}} \wedge \tilde{\gamma}_{\bar{A}}\,, \qquad
\theta^{\alpha } \Xi_{\alpha abc} =\tilde{\lambda}^{\bar{A}} H_{abc\bar{A}} \,, \qquad
\Theta^{\alpha ab} \wedge \rho_{\alpha} = \Lambda^{ab\bar{A}} \wedge \tilde{\gamma}_{\bar{A}}\,, \qquad
\Theta^{\alpha ab} \Xi_{\alpha cde} = \Lambda^{ab\bar{A}} H_{cde\bar{A}}\,, \qquad
\end{equation}
which are a straightforward consequence of the identity (\ref{eq:KappaIdentity}). The purpose of introducing these quantities lies in the fact that they help us eliminate the $\bar{A}$ indices from equations. Note that in both the definitions (\ref{eq:Hdefinitions}) and the identities (\ref{eq:Hidentities}) the indices $\bar{A}$ are summed over on the right-hand sides, while they do not appear at all on the left-hand sides.

It is important to empasize that the arbitrariness of $\kappa^{\alpha\beta}$ in (\ref{eq:ArbitrarinessOfKappaMatrix}) introduces changes into the action. This is due to the fact that the change of variables (\ref{eq:Hdefinitions}) introduces additional variables which do not appear in the original action. The requirement that these additional variables are absent, i.e., that the left-hand sides of identities (\ref{eq:Hidentities}) have the same number of components as the corresponding right-hand sides, reduces the arbitrariness (\ref{eq:ArbitrarinessOfKappaMatrix}) of $\kappa^{\alpha\beta}$ to the following more specific form:
\begin{equation} \label{eq:SmallerArbitrarinessOfKappaMatrix}
\kappa^{\alpha\beta} \to \kappa^{\alpha\beta} + \Big[ \delta^{\alpha}_\gamma - P_\gamma{}^{\alpha}  \Big] A^{\gamma\delta} \Big[ \delta^{\beta}_\delta - P_\delta{}^{\beta}  \Big] \,.
\end{equation}
Note that, although this transformation still allows one to choose $\kappa^{\alpha\beta}$ to be invertible, the action in fact remains invariant with respect to (\ref{eq:SmallerArbitrarinessOfKappaMatrix}), meaning that we can keep working with the same theory. See Appendix \ref{app:d} for a detailed analysis and proof.

Once all these new quantities and notation have been introduced, we can return to the analysis of the action. Given the transformation (\ref{eq:SchematicTransfOfAction}) of the action (\ref{eq:dej}), one can apply the definitions (\ref{eq:Hdefinitions}) and the identities (\ref{eq:Hidentities}) to eliminate all indices $\bar{A},\bar{B}$ and thus rewrite the transformed action $\tilde{S}$ so that it becomes a functional over the reduced configuration space (\ref{eq:ReducedConfSpace}). Putting together the results (\ref{eq:TransfOfTrivialActions}), (\ref{eq:TransfOfHiggsAction}), (\ref{eq:TransfOf3BFAction}) and (\ref{eq:TransfOfScalAction}), we obtain the following form of the full action:
\begin{equation} \label{eq:ElectroweakActionTransformed}
\begin{array}{ccl}
\tilde{S} & = & \ds S_\text{grav} + S_\text{Yang-Mills} + S_\text{CC} + \tilde{S}_\text{Higgs} + \tilde{S}_{3BF} + \tilde{S}_\text{scal} 
  \vphantom{\ds\int} \\
  & & \ds \hphantom{mm} + \int \Theta^{\alpha ab} \wedge \left( \Xi_{\alpha abc} \;\varepsilon^{cdef}e_d\wedge e_e\wedge e_f + \frac{v}{2} \kappa^{-1}_{\alpha\beta} P_\gamma{}^\beta \alpha^{\gamma} \wedge e_a\wedge e_b \right) 
  \vphantom{\ds\int} \\
  & & \ds \hphantom{mmmmm}+ \int \frac{v}{2} \alpha^\alpha P_\alpha{}^\beta \wedge \rho_{\beta} + \theta^{\alpha}\wedge\left(\rho_{\alpha}-\Xi_{\alpha abc}e^a\wedge e^b\wedge e^c\right) 
  \vphantom{\ds\int} \\
  & & \ds \hphantom{mmmmmmmm} + \frac{1}{2} \int  h \, \alpha^\alpha P_\alpha{}^\beta \wedge \left( \rho_{\beta} - \kappa^{-1}_{\beta\gamma} \Theta^{\gamma ab} \wedge e_a\wedge e_b \right) + v \int \rmd \tilde{\gamma}_H \,.
  \vphantom{\ds\int} \\
\end{array}
\end{equation}
This form of the action can now be finally compared to the Proca action (\ref{eq:ProcaAction}). Specifically, the second and third row of (\ref{eq:ElectroweakActionTransformed}) should be compared to the Proca constraint term in the form (\ref{eq:ExtendedProcaConstraintTerm}). The term-by-term comparison gives us the identification of the free parameters in the Proca action as follows:
\begin{equation}
  N_{\alpha\beta} = \frac{v}{2} \kappa^{-1}_{\alpha\gamma} P_\beta{}^\gamma\,, \qquad
  \tilde{N}_\alpha{}^\beta = \frac{v}{2} P_\alpha{}^\beta\,.
\end{equation}
Using these, we can construct the squared-mass matrix (\ref{eq:SquaredMassMatrixDef}) to obtain
\begin{equation}\label{smm}
M^\alpha{}_\beta = \frac{v^2}{4} \left( C^{-1} \right)^{\alpha\gamma} P_\gamma{}^\delta \kappa^{-1}_{\delta\epsilon} P_\beta{}^\epsilon\,, 
\end{equation}
where the coupling constant matrix $C_{\alpha\beta}$ is specified in (\ref{eq:CouplingConstantMatrix}), while the projector and the bilinear form $\kappa^{\alpha\beta}$ are specified in (\ref{eq:ProjectorAndKappaMatrices}). This gives us an explicit form for the squared-mass matrix as:
\begin{equation} \label{eq:squaredMassMatrix}
  M^\alpha{}_\beta = \frac{v^2}{4} \begin{pmatrix}
    g_0^2 & 0 & 0 & -g_0^2 \\
    0 & g_1^2 & 0 & 0 \\
    0 & 0 & g_1^2 & 0 \\
    -g_1^2 & 0 & 0 & g_1^2
\end{pmatrix}\,.
\end{equation}
The physically relevant basis in the Lie algebra $\mathfrak{g}$ is the one in which the above squared-mass matrix is diagonal, and the corresponding eigenvalues are interpreted as squares of masses of gauge vector bosons in that basis. Therefore, we wish to explicitly obtain this basis. Given that the first and last column in (\ref{eq:squaredMassMatrix}) are proportional, the determinant of the squared-mass matrix is zero, meaning that at least one of its eigenvalues is zero. Moreover, the matrix is already in block-diagonal form, with $g_1^2$ being two 1-dimensional blocks, from which one can conclude that two eigenvalues are the same and are equal to $v^2 g_1^2/4$. Finally, from the trace of the matrix one can deduce the fourth eigenvalue, so that the whole set is given as:
\begin{equation} \label{eq:squaredMassEingenvalues}
M_1^2 = 0\,, \qquad M_2^2 = \frac{v^2}{4} g_1^2 \,, \qquad  M_3^2 = \frac{v^2}{4} g_1^2 \,, \qquad M_4^2 = \frac{v^2}{4} (g_0^2 + g_1^2)\,.
\end{equation}
The fact that the eigenvalues $M_2^2$ and $M_3^2$ are equal implies that the eigenbasis is not uniquely determined, and we need some additional input in order to fix it. A natural choice is the eigenbasis of the stabilizer $Q$, introduced in (\ref{eq:StabilizerDef}), since we want to interpret it as the electromagnetic charge, and the value of this charge should be well-defined for each physical state described by our preferred basis. One can reexpress the stabilizer in the matrix form $Q_\alpha{}^\beta$, defined by the action of $Q$ onto the basis vector $\tau_\alpha$:
\begin{equation}
Q \triangleright \tau_\alpha = Q_\alpha{}^\beta \tau_{\beta}\,.
\end{equation}
Using (\ref{eq:StabilizerDef}), one can easily evaluate the components of the matrix $Q_\alpha{}^\beta$ to be
\begin{equation} \label{eq:StabilizerMatrix}
  Q_\alpha{}^\beta = \begin{pmatrix}
    0 & 0 & 0 & 0 \\
    0 & 0 & -1 & 0 \\
    0 & 1 & 0 & 0 \\
    0 & 0 & 0 & 0
\end{pmatrix}\,.
\end{equation}
This matrix has the eigenvalues $(0,i,-i,0)$, so that the electromagnetic charge operator $-iQ$ has the corresponding eigenvalues:
\begin{equation}
q_1 = 0\,, \qquad q_2 = +1\,, \qquad q_3 = -1\,, \qquad q_4 = 0\,.
\end{equation}
The stabilizer also has two equal eigenvalues, so that its eigenbasis is not uniquely determined either. Nevertheless, the squared-mass matrix and the stabilizer matrix mutually commute and therefore share a joint eigenbasis, and this joint eigenbasis is uniquely determined. We can express the new basis in terms of the old basis as follows,
\begin{equation}\label{eq:PhysicalBasisInG}
  \tau_A=\tau_0+\tau_3\,, \qquad
  \tau_+=\frac{\tau_1+i\tau_2}{\sqrt{2}}\,,\qquad
  \tau_-=\frac{\tau_1-i\tau_2}{\sqrt{2}}\,,\qquad
  \tau_Z=-\frac{g_0^2}{g_0^2+g_1^2} \tau_0 + \frac{g_1^2}{g_0^2+g_1^2} \tau_3\,,
\end{equation}
and we can express the components of the connection $1$-form $\alpha = \alpha^\alpha{}_\mu \rmd x^\mu \otimes \tau_\alpha$ in the new basis as:
\begin{equation}
  A_{\mu}= \frac{g_1^2}{g_0^2+g_1^2} \alpha^0{}_{\mu} + \frac{g_0^2}{g_0^2+g_1^2} \alpha^3{}_{\mu}\,,\qquad
  W_{\mu}^{+}=\frac{\alpha^1{}_{\mu}-i \alpha^2{}_{\mu}}{\sqrt{2}}\,,\qquad
  W_{\mu}^{-}=\frac{\alpha^1{}_{\mu}+i\alpha^2{}_{\mu}}{\sqrt{2}}\,,\qquad
  Z_{\mu}=-\alpha^0{}_{\mu}+\alpha^3{}_{\mu}\,.
\end{equation}
Here we have also introduced the traditional notation for the gauge vector bosons. The electromagetic charges of the four bosons are already built into the notation, while their masses can be read from (\ref{eq:squaredMassEingenvalues}):
\begin{equation} \label{eq:MassEingenvalues}
M_A = 0\,, \qquad M_{W^\pm} = \frac{v}{2} g_1 \,, \qquad M_Z = \frac{v}{2} \sqrt{g_0^2 + g_1^2}\,.
\end{equation}
In the new basis, the squared-mass matrix and the stabilizer matrix are diagonal, while the bilinear form $g_{\alpha\beta}$ and the gauge coupling constant bilinear form $C_{\alpha\beta}$ become:
\begin{equation}
g_{\alpha\beta}=\begin{pmatrix}
2 & 0 & 0 & \frac{g_1^2-g_0^2}{g_1^2+g_0^2}\\
0 & 0 & 1 & 0\\
0 & 1 & 0 & 0\\
\frac{g_1^2-g_0^2}{g_1^2+g_0^2} & 0 & 0 & \frac{g_1^4 + g_0^4}{(g_1^2+g_0^2)^2}
\end{pmatrix}\,, \qquad
C_{\alpha\beta}=\begin{pmatrix}
\frac{g_0^2+g_1^2}{g_0^2g_1^2} & 0 & 0 & 0\\
0 & 0 & \frac{1}{g_1^2} & 0\\
0 & \frac{1}{g_1^2} & 0 & 0\\
0 & 0 & 0 & \frac{1}{g_0^2+g_1^2}
\end{pmatrix} \equiv \begin{pmatrix}
\frac{1}{g_A^2} & 0 & 0 & 0\\
0 & 0 & \frac{1}{g_W^2} & 0\\
0 & \frac{1}{g_W^2} & 0 & 0\\
0 & 0 & 0 & \frac{1}{g_Z^2}
\end{pmatrix}\,.
\end{equation}

Returning to the action (\ref{eq:ElectroweakActionTransformed}), we see that the second and third rows represent the appropriate Proca constraint term, which leaves us with the action in the following final form:
\begin{equation} \label{eq:finalFormOfTheAction}
\begin{array}{ccl}
\tilde{S} & = & \ds S_\text{grav} + S_\text{Yang-Mills} + S_\text{CC} + \tilde{S}_\text{Higgs} + \tilde{S}_{3BF} + \tilde{S}_\text{scal} + S_\text{Proca}
  \vphantom{\ds\int} \\
  & & \ds + \frac{1}{2} \int  h \, \alpha^\alpha P_\alpha{}^\beta \wedge \left( \rho_{\beta} - \kappa^{-1}_{\beta\gamma} \Theta^{\gamma ab} \wedge e_a\wedge e_b \right) + v \int \rmd \tilde{\gamma}_H \,.
  \vphantom{\ds\int} \\
\end{array}
\end{equation}
The first row in the action contains terms which describe one real scalar field $h$ (the Higgs field) with mass $m = 2v\sqrt{2\chi}$, and four vector bosons with masses specified in (\ref{eq:MassEingenvalues}), coupled to gravity and to each other. The first term in the second row describes the interaction between the Higgs field and the vector bosons, so that all interactions are equvalent to the interactions of the ordinary electroweak theory. The second term in the second row is a boundary term, and as such it does not contribute to the equations of motion of the theory.

This concludes our analysis of the Higgs mechanism in the context of constrained $3BF$ theory. In short, the result is the same as in the textbook approach to the spontaneous symmetry breaking in electroweak theory. Nevertheless, the technical details that enter the analysis are novel and completely different from the textbook approach, since the $3BF$ formulation of the electroweak action is specified over a different configuration space.

\section{\label{secVI}Conclusions}

\subsection{\label{secVIa}Summary of the results}

Let us summarize the results of the paper. In Section \ref{secII}, we gave a review of the action representing the Standard Model coupled to Einstein-Cartan gravity, within the framework of higher gauge theory. In particular, the action of the model is written as a constrained $3BF$ action, based on a convenient choice of a $3$-group representing the gauge symmetry of the model. Section \ref{secII} also features one Theorem (proved in Appendix \ref{app:a}) which is important for the study of spontaneous symmetry breaking within the higher gauge theory framework, and represents a new result. Section \ref{secIII} was devoted to the study of explicit symmetry breaking of the gauge group of the topological $3BF$ sector, due to the presence of the constraints. Each constraint was studied separately, and we discussed which gauge sector is being broken by which constraint. The results have been summarized in the table. In Section \ref{secIV} we turned our attention to the $3BF$ formulation of the theory for the Proca field coupled to gravity. This was important for the subsequent comparison with the action for the electroweak theory after spontaneous symmetry breaking. Three completely novel and different formulations of the Proca constraint have been discussed, the first for a single Proca field, the second for multiple Proca fields, and the third also for multiple Proca fields in an extended configuration space convenient for comparison with the electroweak model. Finally, in Section \ref{secV} we took up the main task of studying the spontaneous symmetry breaking and the Higgs mechanism for the $3BF$ formulation of the electroweak model. While the Higgs mechanism is conceptually the same as in the ordinary textbook presentations of the electroweak theory, the structure and details of the $3BF$ version of the action are very different from the standard textbook approach, so much that the complete procedure of spontaneous symmetry breaking had to be done anew, with many highly nontrivial details of the calculation. In this sense, the details of the symmetry breaking procedure described in Section \ref{secV} represent one of the main results of the paper.

\subsection{\label{secVIb}Discussion}

Regarding the above results, there are two main comments that need to be addressed. The first comment deals with the question what happens with the structure of the $3$-group as a consequence of the spontaneous symmetry breaking. Namely, the initial $2$-crossed module corresponding to the electroweak theory was based on the following choice of the groups (see equation (\ref{eq:electroweakThreeGroup})):
\begin{equation} \label{eq:initialTwoCrossedModule}
G=SO(3,1)\times SU(2)\times U(1)\,, \qquad H=\mathbb{R}^4\,, \qquad L=\mathbb{C}^4\,.
\end{equation}
However, after the analysis, the resulting action (\ref{eq:finalFormOfTheAction}) does not correspond anymore to this $2$-crossed module. Instead, it is straightforward to see that the final $2$-crossed module is based on the following choice of the groups:
\begin{equation} \label{eq:finalTwoCrossedModule}
G=SO(3,1)\times SU(2) \times U(1)\,, \qquad H=\mathbb{R}^4\,, \qquad L=\realni \,.
\end{equation}
There are two important points to emphasize here. First, the group $L$ has been reduced from the one describing four complex scalar fields to the one describing a single real scalar field. This is a direct consequence of the spontaneous symmetry breaking, in particular of the gauge transformation (\ref{eq:TransformationRuleForPhi}) which was applied to gauge away the fields $\phi_1$, $\phi_2$ and $\phi_4$. The gauge transformation has induced a reduction of the configuration space of the theory, leaving only a single real scalar field $h$ remaining in the action. In turn, the reduction of the configuration space is consistent with the choice (\ref{eq:finalTwoCrossedModule}) of the $2$-crossed module, which corresponds precisely to the topological $3BF$ action (\ref{eq:ActionCalculation2}), obtained by the gauge fixing procedure from the initial $3BF$ action based on the $2$-crossed module (\ref{eq:initialTwoCrossedModule}).

The second point considers the group $G$. Formally, the group $G$ remains the same in both the initial and the final $2$-crossed module. Nevertheless, as we have seen in Section \ref{secIII}, the Proca constraint in fact breaks the $G$ symmetry group, and is the only constraint to do so. Therefore, despite the fact that the topological $3BF$ sector of the initial and final actions shares the same $BF$ term and the same connection 1-form $\alpha$ stemming from the group $G$, the presence of the Proca constraint in the final action in fact breaks the group $G$ down to its subgroup $SO(3,1) \times U(1)$, whereas the initial action did not feature the Proca constraint and the group $G$ was not broken. The end result is that the final action has a broken $G$ symmetry, despite the fact that it is based on the $2$-crossed module (\ref{eq:finalTwoCrossedModule}) featuring the full group $G$. This happens due to the appearance of the Proca constraint during the spontaneous symmetry breaking of the action.

The second comment that needs to be addressed deals with the question of the spontaneous symmetry breaking of the whole Standard Model action (\ref{eq:RealisticAction}). Namely, in Section \ref{secV} we have studied the details of the spontaneous symmetry breaking and the Higgs mechanism on the special case of the electroweak theory, in order to keep the analysis as simple as possible. Nevertheless, it is straightforward to add the remaining three constraints $S_\text{Dirac}$, $S_\text{Yukawa}$, and $S_\text{spin}$, as well as the corresponding $\langle D \wedge {\cal{H}} \rangle_\mathfrak{l}$ term for fermions to the action, and examine the same procedure for the full Standard Model. The resulting action will have the same terms as the action for the electroweak theory, up to terms corresponding to fermions, and up to the overall presence of the color $SU(3)$ gauge symmetry (which remains unbroken and does not play a role in the Higgs mechanism). The $\langle D\wedge {\cal{H}}\rangle_\mathfrak{l}$ term for fermions is equal to
\begin{equation}
\langle D_f\wedge {\cal{H}}_f\rangle_\mathfrak{l}\equiv\bar{\psi}_A(\nablar \gamma)^A-(\bar{\gamma} \nablal)_A\psi^A\,,
\end{equation}
and it transforms trivially under the transformation (\ref{eq:SchematicTransfOfAction}) as well as the $S_\text{Dirac}$ and $S_\text{spin}$ constraints
\begin{equation}
\langle D_f\wedge {\cal{H}}_f\rangle_\mathfrak{l} \;\; \xrightarrow{e^{-\xi}} \;\; \langle \tilde{D}_f\wedge \tilde{\cal{H}}_f\rangle_\mathfrak{l} = \langle D_f\wedge {\cal{H}}_f\rangle_\mathfrak{l} \,, \qquad S_\text{Dirac} \;\; \xrightarrow{e^{-\xi}} \;\; \tilde{S}_\text{Dirac} = S_\text{Dirac} \,, \qquad S_\text{spin} \;\; \xrightarrow{e^{-\xi}} \;\; \tilde{S}_\text{spin} = S_\text{spin} \,.\vphantom{\int}
\end{equation}
The only constraint which does not transform trivially is $S_\text{Yukawa}$, and it splits into two terms:
\begin{eqnarray}
\nonumber
S_\text{Yukawa}=-\int \frac{2}{4!}Y_{ABC}\bar{\psi}^A\psi^B\phi^C\varepsilon_{abcd}e^a\wedge e^b\wedge e^c\wedge e^d \;\; \xrightarrow{e^{-\xi}} \;\; &-&\frac{1}{12}\int vY_{ABH}\,\bar{\psi}^A\psi^B\,\varepsilon_{abcd}\,e^a\wedge e^b\wedge e^c\wedge e^d\\
&-&\frac{1}{12}\int Y_{ABH}\, \bar{\psi}^A\psi^B h \,\varepsilon_{abcd}\, e^a\wedge e^b\wedge e^c\wedge e^d\,,\hphantom{rr}
\end{eqnarray}
where the first term on the right hand side has the form similar to the Dirac mass term (\ref{Dirakova masa}), while the second term is the new Yukawa constraint $\tilde{S}_\text{Yukawa}$, describing the interaction between fermions and the Higgs field $h$. Comparing the first term with the Dirac mass term, we conclude that Yukawa couplings $Y_{ABH}$ are proportional to fermion mass matrix
\begin{equation}
M_{AB}=vY_{ABH}\,,
\end{equation}
which consists of the actual fermion masses and the corresponding mixing angles. The final form of the transformed Yukawa constraint thus becomes:
\begin{equation}
S_\text{Yukawa} \;\; \xrightarrow{e^{-\xi}} \;\; \tilde{S}_\text{Yukawa}-\frac{1}{12}\int M_{AB}\bar{\psi}^A\psi^B\varepsilon_{abcd}e^a\wedge e^b\wedge e^c\wedge e^d=\tilde{S}_\text{Yukawa}+S_\text{Dirac mass}\,.
\end{equation}
Thus, we conclude that the Higgs mechanism described for the electroweak model can be generalized in a straightforward way to the full Standard Model action (\ref{eq:RealisticAction}).

\acknowledgments

The authors would like to thank Tijana Radenkovi\'c and Mihailo \Dj or\dj evi\'c for useful discussions and suggestions.

\medskip

This research was supported by the Ministry of Education, Science and Technological Development of the Republic of Serbia (MPNTR). MV was additionally supported by the Science Fund of the Republic of Serbia, grant number 7745968, project ``Quantum Gravity from Higher Gauge Theory 2021'' -- QGHG-2021. The contents of this publication are the sole responsibility of the authors and can in no way be taken to reflect the views of the Science Fund of the Republic of Serbia.

\appendix

\section{\label{app:a}Proof of the Theorem from the main text}

\textbf{Theorem.} Given a $2$-crossed module $(L\xrightarrow{\delta}H\xrightarrow{\partial}G,\ \triangleright,\ \{\_\, , \_\}_{\rm pf})$ and symmetric, nondegenerate bilinear forms $\langle\_\, ,\_\rangle_{\mathfrak{g}}$, $\langle\_\, ,\_\rangle_{\mathfrak{h}}$  and $\langle\_\, ,\_\rangle_{\mathfrak{l}}$, if the bilinear forms are $G$-invariant then the components of the action $\triangleright_{\alpha\beta\gamma}$, $\triangleright_{\alpha ab}$ and $\triangleright_{\alpha AB}$ are antisymmetric with respect to the second and third index. In addition, there exists a choice of basis in Lie algebras $\mathfrak{g}$, $\mathfrak{h}$ and $\mathfrak{l}$ such that $\triangleright_{\alpha\beta}{}^\gamma$, $\triangleright_{\alpha a}{}^b$ and $\triangleright_{\alpha A}{}^B$ have vanishing diagonal elements with respect to the second and third index, and in this basis the bilinear form is also diagonal.

\medskip

\underline{Proof.}

\medskip

Let us first note that the statement of $G$-invariance of the bilinear form $\langle\_\, ,\_\rangle_{\mathfrak{h}}$ is defined as
\begin{equation}
\langle g\triangleright h_1 , g\triangleright h_2\rangle_{\mathfrak{h}} = \langle h_1, h_2 \rangle_{\mathfrak{h}}\,,
\end{equation}
for all $g \in G$ and all $h_1, h_2 \in \mathfrak{h}$. Expanding $g$, $h_1$ and $h_2$ in appropriate bases, one can easily see that the left-hand side can be rewritten as:
\begin{equation}
\langle g\triangleright h_1 , g\triangleright h_2\rangle_{\mathfrak{h}} = h_1^a h_2^c \Big[ g_{ac} + g^\alpha \big( \triangleright_{\alpha c}{}^d g_{ad} + \triangleright_{\alpha a}{}^d g_{dc} \big) \Big] + \cO(g^2)\,.
\end{equation}
Equating this to the right-hand side, one sees that the right-hand side cancels the first term in the brackets. Then, emloying the symmetry of the bilinear form, the term in the parentheses reduces to:
\begin{equation} \label{eq:AntisymmetryOfTriangle}
\triangleright_{\alpha ca} + \triangleright_{\alpha ac}= 0\,,
\end{equation}
as was stated by the Theorem. The antisymmetry of the remaining two actions $\triangleright_{\alpha\beta\gamma}$ and $\triangleright_{\alpha AB}$ is proved analogously.

In addition, the nondegeneracy of the bilinear form $g_{ab}$ implies that there exists its inverse, denoted $g^{ab}$. Then, contracting (\ref{eq:AntisymmetryOfTriangle}) with $g^{ac}$ one immediately obtains a basis-independent statement that the action is traceless,
\begin{equation}
\triangleright_{\alpha a}{}^a = 0.
\end{equation}
Moreover, one can always choose a basis in a Lie algebra $\mathfrak{h}$ such that the bilinear form $g_{ab}$ and its inverse are diagonal. From the identity
\begin{equation}
\triangleright_{\alpha a}{}^b = \triangleright_{\alpha ac} g^{cb} 
\end{equation}
one can observe that, in this particular basis, $\triangleright_{\alpha a}{}^b$ must be proportional to $\triangleright_{\alpha ab}$ since $g^{cb}=0$ for $c \neq b$. Then, since $\triangleright_{\alpha ab}$ is antisymmetric due to (\ref{eq:AntisymmetryOfTriangle}), it is equal to zero for $a=b$, which implies that $\triangleright_{\alpha a}{}^b$ is also zero in that case. In other words,  $\triangleright_{\alpha a}{}^b$ has vanishing diagonal elements with respect to second and third index, as was stated by the Theorem. The same property for the remaining two actions $\triangleright_{\alpha\beta}{}^\gamma$ and $\triangleright_{\alpha A}{}^B$ is proved analogously.

\section{\label{app:d}Arbitrariness of the $\kappa$-matrix}

In Section \ref{secV} we introduced the bilinear form $\kappa^{\alpha\beta}$ via (\ref{eq:KappaIdentity}) as well as the new variables (\ref{eq:Hdefinitions}) which satisfy the list of identities (\ref{eq:Hidentities}). In order to maintain one-to-one correspondence between the old and the new variables, in these identities the number of independent variables on the right hand side has to be equal to the number of the independent variables on the left hand side. Moreover, none of these new variables (\ref{eq:Hdefinitions}) should be explicitly multiplied by zero during the construction of the identities (\ref{eq:Hidentities}). These requirements have nontrivial consequences on arbitrariness of the choice of the bilinear form $\kappa^{\alpha\beta}$. For instance, let us consider the first identity from (\ref{eq:Hidentities}),
\begin{equation}\label{pomocniidentitetD}
\theta^\alpha\wedge\rho_\alpha=\lambda^{\Bar{A}}\wedge\gamma_{\Bar{A}}.
\end{equation}
From the definition of the variable $\rho_\alpha$ in (\ref{eq:Hdefinitions}) we conclude that the action of the projector (\ref{eq:ProjectorDef}) does not change $\rho_\alpha$
\begin{equation}
\rho_\alpha=P_{\alpha}{}^{\beta}\rho_\beta.
\end{equation}
This implies that this projector does not change the left hand side of identity (\ref{pomocniidentitetD}) and it holds that
\begin{equation}
\theta^\alpha\wedge\rho_\alpha=\theta^\alpha\wedge P_{\alpha}{}^{\beta}\rho_\beta.
\end{equation}
Using the requirement that none of the variables should be multiplied by zero during the construction of the identity (\ref{pomocniidentitetD}), we conclude that the action of projector must also leave the variable $\theta^\alpha$ invariant. Using the definition of the $\theta^\alpha$ variable we obtain a nontrivial condition on $\kappa^{\alpha\beta}$:
\begin{equation}
  \theta^\alpha=\theta^{\beta}P_{\beta}{}^{\alpha} \qquad \Rightarrow \qquad -2 \kappa^{\alpha\gamma} \triangleright_{\gamma}{}^H{}_{\bar{A}}\lambda^{\bar{A}} \equiv -2 \kappa^{\alpha\gamma} P_{\gamma}{}^{\delta}\triangleright_{\delta}{}^H{}_{\bar{A}}\lambda^{\bar{A}}=-2 P_{\beta}{}^{\alpha}\kappa^{\beta\gamma} P_{\gamma}{}^{\delta}\triangleright_{\delta}{}^H{}_{\bar{A}}\lambda^{\bar{A}}\,,
\end{equation}
which due to arbitrariness of the field $\lambda^{\Bar{A}}$ implies that
\begin{equation}\label{Kapa_uslovD}
\kappa^{\alpha\gamma} P_{\gamma}{}^{\delta}=P_{\beta}{}^{\alpha}\kappa^{\beta\gamma} P_{\gamma}{}^{\delta}\,.
\end{equation}
Using the fact that $\kappa^{\alpha\beta}$ is symmetric, by transposing (\ref{Kapa_uslovD}) we obtain that the projector and $\kappa^{\alpha\beta}$ commute
\begin{equation}\label{komutatorPK}
\kappa^{\alpha\gamma} P_{\gamma}{}^{\delta}=P_{\gamma}{}^{\alpha}\kappa^{\gamma\delta} \,.
\end{equation}
Now, let us consider how this new restriction on $\kappa^{\alpha\beta}$ reduces its arbitrariness. Combining (\ref{eq:ArbitrarinessOfKappaMatrix}) and (\ref{komutatorPK}) we obtain
\begin{equation}
P_{\gamma}{}^{\alpha}A^{\gamma\delta}\Big[\delta_{\delta}^{\beta}-P_{\delta}{}^{\beta}\Big]=\Big[\delta_{\gamma}^{\alpha}-P_{\gamma}{}^{\alpha}\Big]A^{\gamma\delta}P_{\delta}{}^{\beta}\,,
\end{equation}
which is satisfied only if
\begin{equation}
A^{\alpha\delta}\Big[\delta_{\delta}^{\beta}-P_{\delta}{}^{\beta}\Big]=\Big[\delta_{\gamma}^{\alpha}-P_{\gamma}{}^{\alpha}\Big]A^{\gamma\beta}=\Big[\delta_{\gamma}^{\alpha}-P_{\gamma}{}^{\alpha}\Big]A^{\gamma\delta}\Big[\delta_{\delta}^{\beta}-P_{\delta}{}^{\beta}\Big]\,.
\end{equation}
Since the dimension of the subspace of the orthogonal projector is equal to one, the arbitrariness in the choice of the bilinear form $\kappa^{\alpha\beta}$ is reduced to a single free parameter, whose form is given in (\ref{eq:SmallerArbitrarinessOfKappaMatrix}):
\begin{equation}
\kappa^{\alpha\beta} \to \kappa^{\alpha\beta} + \Big[ \delta^{\alpha}_\gamma - P_\gamma{}^{\alpha}  \Big] A^{\gamma\delta} \Big[ \delta^{\beta}_\delta - P_\delta{}^{\beta}  \Big] \,.
\end{equation}
The arbitrariness of this parameter guarantees the existence of the inverse bilinear form $\kappa^{-1}_{\alpha\beta}$, which also commutes with the projector as a consequence of (\ref{komutatorPK}).

Next, we turn to the effect of this arbitrariness onto the action (\ref{eq:ElectroweakActionTransformed}). Using the fact that the inverse bilinear form acts on variables (\ref{eq:Hdefinitions}) in (\ref{eq:ElectroweakActionTransformed}), and variables (\ref{eq:Hdefinitions}) are invariant under projector action, the action (\ref{eq:ElectroweakActionTransformed}) of the theory, and squared-mass matrix (\ref{smm}) depend only on the projection of inverse bilinear form $\kappa^{-1}_{\alpha\beta}P_{\gamma}{}^{\beta}$. The arbitrariness of the inverse bilinear form $\kappa^{-1}_{\alpha\beta}$ can be expressed as a power series in terms of the arbitrary bilinear form $A^{\gamma\delta}$, as
\begin{equation}
\kappa^{-1}_{\alpha\beta} \to \kappa^{-1}_{\alpha\gamma}\sum_{n=0}^{\infty}\left[(-1)^n\left(\left[ \delta - P\right] A \left[ \delta - P\right]\kappa^{-1}\right)^{n}\right]^{\gamma}{}_\beta \,,
\end{equation}
from where one can obtain that the projection of the inverse bilinear form does not depend on choice of the arbitrary bilinear form $A^{\gamma\delta}$. This in turn implies that the action (\ref{eq:ElectroweakActionTransformed}) and the squared-mass matrix (\ref{smm}) are uniquely defined.


\begin{thebibliography}{99}

\bibitem{Polchinski1}
J. Polchinski,
{\it String Theory Vol. I: An Introduction to the Bosonic String}, Cambridge University Press, Cambridge (1998).

\bibitem{Polchinski2}
J. Polchinski,
{\it String Theory Vol. II: Superstring Theory and Beyond}, Cambridge University Press, Cambridge (1998).

\bibitem{Rovelli2004}
C. Rovelli,
{\it Quantum Gravity}, Cambridge University Press, Cambridge (2004).

\bibitem{Thiemann}
T. Thiemann,
{\it Modern Canonical Quantum General Relativity}, Cambridge University Press, Cambridge (2007).

\bibitem{RovelliVidotto2014}
C. Rovelli and F. Vidotto,
{\it Covariant Loop Quantum Gravity},  Cambridge University Press, Cambridge (2014).

\bibitem{SpinfoamFermions}
E. Bianchi, M. Han, E. Magliaro, C. Perini, C. Rovelli and W. Wieland,
``Spinfoam fermions'',
{\it Class. Quant. Grav.} {\bf 30}, 235023 (2013),
\texttt{arXiv:1012.4719}.
  
\bibitem{VojinovicCosineProblem}
M. Vojinovi\'c,
``Cosine problem in EPRL/FK spin foam model'',
{\it Gen. Relativ. Gravit.} {\bf 46}, 1616 (2014),
\texttt{arXiv:1307.5352}.

\bibitem{MikovicVojinovicBook}
A. Mikovi\'c and M. Vojinovi\'c,
{\it State-Sum Models of Piecewise Linear Quantum Gravity}, World Scientific, Singapore (2023).

\bibitem{BaezHuerta}
J. C. Baez and J. Huerta,
``An Invitation to Higher Gauge Theory'',
{\it Gen. Relativ. Gravit.} {\bf 43}, 2335 (2011),
{\tt arXiv:1003.4485}.

\bibitem{Crane2003}
L. Crane and M. D. Sheppeard, 
``2-Categorical Poincar\' e representations and state sum applications'',
{\tt arXiv:math/0306440}.

\bibitem{MikovicVojinovic2012}
A. Mikovi\'c and M. Vojinovi\'c,
``Poincar\'e 2-group and quantum gravity'',
{\it Class. Quant. Grav.} {\bf 29}, 165003 (2012),
{\tt arXiv:1110.4694}.
  
\bibitem{Li2019}
Z. Li,
``A global geometric approach to parallel transport of strings in gauge theory'',
\texttt{arxiv:1910.14230}.

\bibitem{SaemannWolf2014}
C, Saemann and M. Wolf,
``Non-abelian tensor multiplet equations from twistor space'',
{\it Commun. Math. Phys.} {\bf 328}, 527 (2014),
\texttt{arxiv:1205.3108}.

\bibitem{JurcoEtal2005}
B. Jurco, C. Saemann and M. Wolf,
``Semistrict Higher Gauge Theory'',
{\it JHEP} {\bf 87} (2005),
\texttt{arxiv:1403.7185}.
  
\bibitem{HNY2020}
Y. Hidaka, M. Nitta and R. Yokokura,
``Higher-form symmetries and 3-group in axion electrodynamics'',
{\it Phys. Lett. B} \textbf{808}, 135672 (2020),
\texttt{arxiv:2006.12532}.
  
\bibitem{HNY2021}
Y. Hidaka, M. Nitta and R. Yokokura,
``Global 3-group symmetry and 't Hooft anomalies in axion electrodynamics'',
{\it JHEP} {\bf 1}, 173 (2021),
\texttt{arxiv:2009.14368}.

\bibitem{SaemannWolf2014b}
C. Saemann and M. Wolf,
``Six-Dimensional Superconformal Field Theories from Principal 3-Bundles over Twistor Space'',
{\it Lett. Math. Phys.} {\bf 104}, 1147 (2014),
\texttt{arxiv:1305.4870}.

\bibitem{SWY2021a}
D. Song, K. Wu and J. Yang,
``3-form Yang-Mills based on 2-crossed modules'',
{\it Jour. Geom. Phys.} {\bf 178}, 104537 (2022),
\texttt{arxiv:2108.12852}.

\bibitem{SWY2021b}
D. Song, K. Wu and J. Yang,
``Higher form Yang-Mills as higher BFYM theories'',
{\it Eur. Phys. Jour. C} {\bf 82}, 1034 (2022),
\texttt{arxiv:2109.13443}.
  
\bibitem{HNY2021b}
Y. Hidaka, M. Nitta and R. Yokokura,
``Topological axion electrodynamics and 4-group symmetry'',
{\it  Phys. Lett. B} {\bf 823}, 136762 (2021),
\texttt{arxiv:2107.08753}.

\bibitem{HNY2021c}
Y. Hidaka, M. Nitta and R. Yokokura,
``Global 4-group symmetry and 't Hooft anomalies in topological axion electrodynamics'',
{\it Prog. Theor. Exper. Phys.} {\bf 2022}, 04A109 (2022),
\texttt{arxiv:2108.12564}.
  
\bibitem{JurcoSaemannWolf2016}
B. Jurco, C. Saemann and M. Wolf,
``Higher Groupoid Bundles, Higher Spaces, and Self-Dual Tensor Field Equations'',
{\it Fortsch. Phys.} {\bf 64}, 674 (2016),
\texttt{arxiv:1604.01639}.

\bibitem{SaemannWolf2017}
C. Saemann and M. Wolf,
``Supersymmetric Yang-Mills Theory as Higher Chern-Simons Theory'',
{\it JHEP} {\bf 07}, 111 (2017),
\texttt{arxiv:1702.04160}.

\bibitem{JurcoEtal2019b}
B. Jurco, T. Macrelli, L. Raspollini, C. Saemann and M. Wolf,
``L-infinity Algebras, the BV Formalism, and Classical Fields'',
{\it Fortsch. Phys.} {\bf 67}, 1910025 (2019),
\texttt{arxiv:1903.02887}.

\bibitem{Radenkovic2019}
T. Radenkovi\' c and M. Vojinovi\' c,
``Higher Gauge Theories Based on 3-groups'',
{\it JHEP} {\bf 10}, 222 (2019),
\texttt{arXiv:1904.07566}.

\bibitem{Radenkovic2022b} 
T. Radenkovi\'c and M. Vojinovi\'c,
``Topological invariant of $4$-manifolds based on a $3$-group'',
{\it JHEP} {\bf 7}, 105 (2022),
{\tt arXiv:2201.02572.}

\bibitem{BV1}
\red{I. A. Batalin and G. A. Vilkovisky,}
\red{``Gauge algebra and quantization'',}
\red{{\it Phys. Lett. B} {\bf 102}, 27 (1981).}

\bibitem{BV2}
\red{I. A. Batalin and G. A. Vilkovisky,}
\red{``Feynman rules for reducible gauge theories'',}
\red{{\it Phys. Lett. B} {\bf 120}, 166 (1983).}
  
\bibitem{BV3}
\red{I. A. Batalin and G. A. Vilkovisky,}
\red{``Quantization of gauge theories with linearly dependent generators'',}
\red{{\it Phys. Rev. D} {\bf 28}, 2567 (1983), errata: {\it Phys. Rev. D} {\bf 30}, 508 (1984).}
  
\bibitem{BV4}
\red{I. A. Batalin and G. A. Vilkovisky,}
\red{``Closure of the gauge algebra, generalized lie equations and Feynman rules'',}
\red{{\it Nucl. Phys.} {\bf B234}, 106 (1984).}
  
\bibitem{BV5}
\red{I. A. Batalin and G. A. Vilkovisky,}
\red{``Existence theorem for gauge algebra'',}
\red{{\it Jour. Math. Phys.} {\bf 26}, 172 (1985).}
  
\bibitem{hep-th/9502010}
\red{M. Alexandrov, A. Schwarz, O. Zaboronsky, and M. Kontsevich,}
\red{``The Geometry of the Master Equation and Topological Quantum Field Theory'',}
\red{{\it Int. Jour. Mod. Phys. A} {\bf 12}, {7} (1997),}
\red{{\tt arXiv:hep-th/9502010}.}

\bibitem{MaximParent}
\red{M. Grigoriev,}
\red{``Parent formulation at the Lagrangian level'',}
\red{{\it JHEP} {\bf 2011}, 61 (2011),}
\red{{\tt arXiv:1012.1903}.}

\bibitem{MaximPresymplectic}
\red{K. B. Alkalaev, M. Grigoriev,}
\red{``Frame-like Lagrangians and presymplectic AKSZ-type sigma models'',}
\red{{\it Int. Jour. Mod. Phys. A}, {\bf 29}, 18 (2014),}
\red{{\tt arXiv:1312.5296}.}

\bibitem{MaximPresymplecticMinimal}
\red{I. Dneprov, M. Grigoriev, V. Gritzaenko,}
\red{``Presymplectic minimal models of local gauge theories'',}
\red{{\tt arXiv:2402.03240}.}

\bibitem{Alberto(Maxim)}
\red{A. S. Cattaneo, L. Menger, M. Schiavina,}
\red{``Gravity with torsion as deformed BF theory'',}
\red{{\it Class. Quant. Grav.} {\bf 41}, 155001 (2024),}
\red{{\tt arXiv:2310.01877}.}

\bibitem{AKSZ-Manin(Maxim)}
\red{L. Borsten, D. Kanakaris, H. Kim,}
\red{``Gravity from AKSZ-Manin theories in two, three, and four dimensions'',}
\red{{\tt arXiv:2410.10755}.}

\bibitem{Jan(Maxim)}
\red{J. Pulmann, P. \v Severa, F. Valach,}
\red{``A non-abelian duality for (higher) gauge theories'',}
\red{{\it Adv. Theor. Math. Phys.}, {\bf 25}, 1 (2021),}
\red{{\tt arXiv:1909.06151}.}

\bibitem{Plebanski}
\red{J. F. Plebanski,}
\red{``On the separation of Einsteinian substructures'',}
\red{{\it Jour. Math. Phys.} {\bf 18}, 2511 (1977).}

\bibitem{baez2000}
J. C. Baez, 
{``An Introduction to Spin Foam Models of Quantum Gravity and BF Theory''},
{\it Lect. Notes Phys.} {\bf 543}, {25} (2000),
{\tt arXiv:gr-qc/9905087}.

\bibitem{BFgravity2016}
M. Celada, D. Gonz\' alez and M. Montesinos,
{``BF gravity''},
{\it Class. Quant. Grav.} {\bf 33}, 213001 {(2016)},
{\tt arXiv:1610.02020}.

\bibitem{Baez1996}
\red{J. Baez,}
\red{``4-Dimensional BF Theory as a Topological Quantum Field Theory'',}
\red{{\it Lett. Math. Phys.} {\bf 38}, 129 (1996),}
\red{\texttt{arxiv:q-alg/9507006}.}

\bibitem{GirelliPfeifferPopescu2008}
F. Girelli, H. Pfeiffer and E. M. Popescu,
{``Topological higher gauge theory -- from BF to BFCG theory'',}
{\it Jour. Math. Phys.} {\bf 49}, 032503 (2008),
{\tt arXiv:0708.3051}.

\bibitem{FariaMartinsMikovic2011}
J. F. Martins and A. Mikovi\'c,
{``Lie crossed modules and gauge-invariant actions for 2-BF theories'',}
{\it Adv. Theor. Math. Phys.} {\bf 15}, 1059 (2011),
{\tt arXiv:1006.0903}.

\bibitem{MikovicOliveira2014}
A. Mikovi\' c and M. A. Oliveira,
{``Canonical formulation of Poincar\'e BFCG theory and its quantization''},
{\it Gen. Relativ. Gravit.} {\bf 47}, 58 (2015),
{\tt arXiv:1409.3751.}

\bibitem{Mikovic2015}
A. Mikovi\' c, M. A. Oliveira and M. Vojinovi\' c,
{``Hamiltonian analysis of the $BFCG$ theory for the Poincar\'e $2$-group''},
{\it Class. Quant. Grav.} {\bf 33}, 065007 (2016),
{\tt arxiv:1508.05635}.

\bibitem{MOV2016}
A. Mikovi\' c, M. A. Oliveira and M. Vojinovi\'c,
``Hamiltonian analysis of the BFCG theory for a generic Lie 2-group'',
{\it Adv. Theor. Math. Phys.} {\bf 26}, 3783 (2022),
{\tt arXiv:1610.09621}.

\bibitem{MOV2019}
A. Mikovi\' c, M. A. Oliveira and M. Vojinovi\'c,
{``Hamiltonian analysis of the BFCG formulation of General Relativity'',}
{\it Class. Quant. Grav.} {\bf 36}, 015005 (2019),
{\tt arXiv:1807.06354.}

\bibitem{Asante2020}
S. K. Asante, B. Dittrich, F. Girelli, A. Riello and P. Tsimiklis,
``Quantum geometry from higher gauge theory'',
{\it Class. Quant. Grav.} {\bf 37}, 205001 (2020),
\texttt{arxiv:1908.05970}.

\bibitem{Girelli2021}
F. Girelli, M. Laudonio and P. Tsimiklis,
``Polyhedron phase space using 2-groups: kappa-Poincare as a Poisson 2-group'',
\texttt{arxiv:2105.10616}.
  
\bibitem{MV2020}
A. Mikovi\'c and M. Vojinovi\'c,
{``Standard Model and 4-groups''},
{\it Europhys. Lett.} {\bf 133}, 61001 (2021),
{\tt arXiv:2008.06354.}

\bibitem{Porter}
\red{T. Porter,}
\red{``Topological quantum field theories from homotopy n-types'',}
\red{{\it J. London Math. Soc.} {\bf 58}, 723 (1998),}
\red{\texttt{MR 1678163}.}

\bibitem{Radenkovic2024}
\red{T. Radenkovi\'c, M. Vojinovi\'c,}
\red{``The 3BF theory as a TQFT'',}
\red{{\tt arXiv:2412.21032}.}

\bibitem{BlagojevicBook}
\red{M. Blagojevi\'c,}
\red{{\it Gravitation and Gauge Symmetries},}
\red{Institute of Physics Publishing, Bristol (2002).}

\bibitem{Radenkovic2020} 
T. Radenkovi\'c and M. Vojinovi\'c,
{``Hamiltonian Analysis for the Scalar Electrodynamics as $3BF$ Theory'',}
{\it Symmetry} {\bf 12}, 620 (2020),
{\tt arXiv:2004.06901.}

\bibitem{Radenkovic2022a} 
T. Radenkovi\'c and M. Vojinovi\'c,
{``Gauge symmetry of the $3BF$ theory for a generic Lie $3$-group'',}
{\it Class. Quant. Grav.} {\bf 39}, 135009 (2022),
{\tt arXiv:2101.04049.}

\bibitem{martins2011}
J. F. Martins and R. Picken,
{``The fundamental Gray 3-groupoid of a smooth manifold and local 3-dimensional holonomy based on a 2-crossed module'',}
{\it Differ. Geom. Appl. Jour.} {\bf 29}, 179-206 {(2011)},
{\tt arXiv:0907.2566.} 


\bibitem{Wang2014}
W. Wang,
{``On 3-gauge transformations, 3-curvatures and Gray-categories'',}
{\it Jour. Math. Phys.} {\bf 55}, 043506 (2014).

\bibitem{BaezDolan1995}
\red{J. Baez and J. Dolan,}
\red{``Higher dimensional algebra and Topological Quantum Field Theory'',}
\red{{\it Jour. Math. Phys.} {\bf 36}, 6073 (1995),}
\red{\texttt{arxiv:q-alg/9503002}.}

\bibitem{Conduche1984}
\red{D. Conduch\'e,}
\red{``Modules crois\'es g\'en\'eralis\'es de longueur 2'',}
\red{Proceedings of the Luminy conference on algebraic K-theory (Luminy, 1983),}
\red{{\it J. PureAppl. Algebra} {\bf 34}, 155 (1984).}

\bibitem{Yetter1993}
\red{D. N. Yetter,}
\red{``TQFT's from homotopy 2-types'',}
\red{{\it J. Knot Theory Ramifications} {\bf 2}, 113 (1993).}

\bibitem{GirreliPfeiffer2004}
\red{F. Girelli and H. Pfeiffer,}
\red{``Higher gauge theory --- differential versus integral formulation'',}
\red{{\it Jour. Math. Phys.} {\bf 45}, 3949 (2004),}
\red{\texttt{arxiv:hep-th/0309173}.}

\bibitem{Berends1985}
\red{F. A. Berends, G. J. H. Burgers and H. van Dam,}
\red{``On the theoretical problems in constructing interactions involving higher spin massless particles'',}
\red{{\it Nucl. Phys.} {\bf B260}, 295 (1985).}

\bibitem{LadaStasheff1993}
\red{T. Lada and J. Stasheff,}
\red{``Introduction to sh Lie algebras for physicists'',}
\red{{\it Int. J. Theor. Phys.} {\bf 32}, 1087 (1993),}
\red{\texttt{arXiv:hep-th/9209099}.}

\bibitem{Stasheff1998}
\red{J. Stasheff,}
\red{``The (secret?) homological algebra of the Batalin-Vilkovisky approach'',}
\red{{\it Contemp. Math.} {\bf 219}, 195 (1998),}
\red{\texttt{arXiv:hep-th/9712157}.}

\bibitem{HohmZwiebach2017}
\red{O. Hohm and B. Zwiebach,}
\red{``L-infinity algebras and field theory'',}
\red{{\it Fortsch. Phys.} {\bf 65}, 1700014 (2017),}
\red{\texttt{arXiv:1701.08824}.}

\bibitem{Djordjevic2023}
M. \Dj or\dj evi\'c, T. Radenkovi\'c, P. Stipsi\'c, and M. Vojinovi\'c,
{``Henneaux-Teitelboim gauge symmetry and its applications to higher gauge theories''},
{\it Universe} {\bf 9}, 281 (2023),
{\tt arXiv:2305.00117.}

\end{thebibliography}
\end{document}